\documentclass[showpacs,floatfix,superscriptaddress]{revtex4}
\usepackage{graphicx,amsfonts,amssymb,amsmath, hyperref}

\def\nbZ{{\mathchoice {\hbox{$\sf\textstyle Z\kern-0.4em Z$}} 
{\hbox{$\sf\textstyle   Z\kern-0.4em    Z$}}   {\hbox{$\sf\scriptstyle
Z\kern-0.3em Z$}} {\hbox{$\sf\scriptscriptstyle Z\kern-0.2em Z$}}}}

\usepackage{latexsym,amsmath,tabularx,amssymb,bm}
\usepackage{graphicx}

\newcommand{\bea}{\begin{eqnarray}}
\newcommand{\eea}{\end{eqnarray}}

\newcommand{\be}{\begin{equation}}
\newcommand{\ee}{\end{equation}}

\usepackage[applemac]{inputenc}

\def\simge{\mathrel{%
   \rlap{\raise 0.511ex \hbox{$>$}}{\lower 0.511ex \hbox{$\sim$}}}}
\def\simle{\mathrel{
   \rlap{\raise 0.511ex \hbox{$<$}}{\lower 0.511ex \hbox{$\sim$}}}}

\def\simle{\mathrel{
   \rlap{\raise 0.511ex \hbox{$<$}}{\lower 0.511ex \hbox{$\sim$}}}}

\def\simge{\mathrel{%
    \rlap{\raise 0.511ex \hbox{$>$}}{\lower 0.511ex \hbox{$\sim$}}}}

\begin{document}

\title{About the relevance of the   fixed dimension perturbative approach to frustrated 
magnets in two and three dimensions}

\author{B. Delamotte}
\affiliation{LPTMC, CNRS-UMR 7600, Universit\'e Pierre et Marie Curie, 75252 Paris Cedex 05, France}

\author{M. Dudka}
\affiliation{Institute  for Condensed   Matter  Physics,  National  Acad.  Sci.   of
Ukraine, UA--79011 Lviv, Ukraine}

\author{Yu. Holovatch}
\affiliation{Institute  for Condensed   Matter  Physics,  National  Acad.  Sci.   of
Ukraine, UA--79011 Lviv, Ukraine}
\affiliation{Institut   f\"ur  Theoretische  Physik, Johannes Kepler  Universit\"at
Linz, A-4040 Linz, Austria}

\author{D. Mouhanna}
\affiliation{LPTMC, CNRS-UMR 7600, Universit\'e Pierre et Marie Curie, 75252 Paris Cedex 05, France}

\begin{abstract}

We show that the critical behaviour of two- and three-dimensional frustrated magnets cannot reliably  be described
from the known  five- and six-loops perturbative renormalization group results. Our conclusions are
based on a careful re-analysis of the resummed perturbative series obtained within the  zero momentum massive scheme. In  three dimensions, the critical exponents for XY and Heisenberg spins display strong dependences
on the parameters of the resummation procedure and on the loop order. This behaviour  strongly suggests  that the fixed points found are in fact  spurious. In two dimensions, we find, as in the $O(N)$ case, that there is apparent convergence of the critical exponents but towards erroneous values.
As a consequence,  the interesting question of the description of the crossover/transition induced by $Z_2$ topological defects in two-dimensional frustrated Heisenberg spins remains open.

\end{abstract}


\maketitle

\vskip 1cm

\section{Introduction}

After more than thirty years of intensive studies, the critical behaviour of frustrated magnets is still  controversial (see \cite{delamotte03} and references therein).  At the  root of the problem is the competition between the interactions among neighboring spins that gives rise to a canted ground state,   and thus, to a symmetry breaking scheme where the rotational group is fully broken.  This is for instance the case in the paradigmatic example  of frustrated magnets,  the Stacked Triangular Antiferromagnets  (STA),  where the three spins on an elementary cell  display   a planar 120$^{\circ}$ structure in the ground state.  As a consequence  the order parameter  is a matrix 
instead of a vector (a $SO(3)$  matrix for Heisenberg spins  and  a $2\times 2$  matrix   for XY spins) and  the critical properties  are therefore entirely different  from those of unfrustrated systems. 

 For instance in  dimension $d=2$,  the first  homotopy group of $SO(3)$ being non trivial -- $\Pi_1(SO(3))=\nbZ_2$ --  one expects  for Heisenberg spins a deconfinement of topological excitations \cite{kawamura84}   that could give   rise to a Kosterlitz-Thouless(KT)-like transition \cite{kosterlitz73,kosterlitz74} or, at least, to a crossover behaviour -- see below.   Numerous  experimental \cite{olariu06,yamaguchi08,yaouanc08,takeya08,hemmida09} and numerical   \cite{kawamura84,southern93,wintel94,southern95,wintel95,caffarel01,messio08,kawamura10,okubo10}  studies have indeed shown indications of   a nontrivial phenomenon  occuring at finite temperature.   For   XY spins   the order parameter space is  given by $SO(2)\times \nbZ_2$.    In this case coexist Ising degrees of freedom,    topological  excitations   -- $\Pi_1(SO(2))=\nbZ$ --  and spin-waves.  A very debated  issue  has been   the nature of the phase transition(s) occuring in this  system  as the temperature is varied: either two separate Ising and KT  transitions or a unique one (see \cite{hasenbusch05} and references therein). 
 
  In   $d=3$  the question of the criticality has been extremely controversial (see \cite{delamotte03}).   On the one hand, many experiments  display   scaling  behaviours  for  XY  and Heisenberg spins with critical  exponents differing from those of the $O(N)$ universality class (see \cite{delamotte03} for a review).  On the other hand, other experiments as well as extensive  Monte Carlo simulations  performed on STA or on  similar  models   have exhibited  weak first order behaviour \cite{diep89,loison98,loison00b, itakura03,peles04,bekhechi06,quirion06,zelli07,thanhngo08}.
 Two main explanations have been proposed to  describe  these contradicting results.

The first one is based  on a  perturbative  renormalization group (RG) approach performed at  fixed dimension (FD) either within the  minimal-substraction ($\overline{\hbox{MS}}$)  scheme without  $\epsilon$-expansion \cite{calabrese04} or within the zero momentum  massive scheme \cite{antonenko94,pelissetto01a,calabrese02,calabrese03b},  at five- and six-loop order respectively. Within these approaches,  stable RG fixed points   were found   for $N=2$ and $N=3$  leading to the prediction of  second order
 phase transitions  in $d=3$ for  frustrated magnets.  Note that, within these FD approaches, one also finds  a fixed point in $d=2$ with nontrivial critical exponents in the $N=2$ and  $N=3$ cases  \cite{calabrese01,calabrese02,calabrese03}. This  fact  has led  to the  hypothesis of a Kosterlitz-Thouless-like behaviour induced by $\nbZ_2$ topological defects for Heisenberg spins  \cite{calabrese01}.

The second explanation  is based on both  the  $\epsilon=4-d$ (or pseudo-$\epsilon$)-expansion \cite{antonenko95,holovatch04,calabrese03c} and   the nonperturbative renormalization group (NPRG) approaches \cite{tissier00,tissier00b,tissier01, tissier01c, delamotte03}. In these approaches, one finds  that there exists, within the $(d,N)$ plane,  a line $N_c(d)$ that separates a second order region  for $N>N_c$  from a first order region for $N<N_c$. Within  the   $\epsilon$-expansion,  one finds $N_c(d=3)\simeq 5.3$ \cite{calabrese03c}, pseudo-$\epsilon$ expansion gives  $N_c(d=3)\simeq 6.23$ \cite{holovatch04} and within the NPRG approach $N_c(d=3)\simeq 5.1$  \cite{delamotte03} so that the transition for   $N=2$  and  $N=3$   are  predicted to be of the first order. A thorough  analysis performed within  the NPRG approach \cite{tissier00,tissier01,delamotte03} has shown that even if there is, strictly speaking,  no fixed point  below  $N_c(d=3)$ the   RG flow is  very slow for $N=2$ and 3 in a whole region of the coupling constant space so  that  there is  pseudo-scaling  without universality on a large range of temperature, in agreement with  the numerical and experimental data.

Although  the NPRG  approach  very likely  explains  the whole body of known data (see \cite{delamotte03} for details),  several  points   of the physics of frustrated magnets remain controversial.    The first one is  that, although the occurence of  (weak) first order transitions is by now well  established in several three-dimensional systems   \cite{diep89,loison98,loison00b, itakura03,peles04,bekhechi06,quirion06,zelli07,thanhngo08}  this  does not  imply  that all  systems, sharing the  same order parameter and the same symmetries,  undergo  first  order phase transitions   (see for instance \cite{calabrese04} for a recent numerical computation where a second order phase transition is observed).  In other words, the very existence of a parameter domain where the frustrated systems would undergo second order transitions is still  debated. A second point  that should be understood  is the origin of the discrepancy between the  two scenarii above and,  in particular, between  the results obtained within the different  perturbative schemes:  $\epsilon$ (or pseudo-$\epsilon$)-expansion  on the one hand and  the FD  approaches on the other hand.  A last and important question is the nature of the transition that occurs  in $d=2$ for Heisenberg spins:  phase transition or simple cross-over behaviour  between a low-temperature -- spin-wave -- phase and a high-temperature phase with both spin-waves and   vortices ?

 It is clear that answering to these  questions amounts to answering to the  question of the existence of a  genuine attractive  fixed point in the RG flow  of frustrated magnets in $d=3$ and $2$ for  $N=2$ and 3. In the perturbative framework, this is essentially equivalent  to proving (or disproving) the reliability   and convergence properties  of the resummation procedures  necessary to obtain sensible results out of  the perturbative series.  This  work has been  initiated in $d=3$ in a previous publication where the  five-loop  perturbative series obtained in the  $\overline{\hbox{MS}}$  scheme {\sl without $\epsilon$-expansion}       have been carefully reexamined \cite{delamotte08}.  Studying (i) the convergence properties of the critical exponents  with the order $L$  of the expansion (number of loops) and with respect to  the variations of the parameters  involved in the resummation procedure,   (ii)   the  properties of the fixed point coordinates   ($u^*(d,N),v^*(d,N)$)  considered  as functions of $d$ and $N$  and (iii)  the continuation of the fixed point  found  in  $d=3$  for $N=2$ and $N=3$ up to $d=4$,   the authors of  \cite{delamotte08}  have   provided  strong arguments  in favor of  the spurious character of  the fixed points  found in $d=3$, {\it i.e.}  that they are  artefacts  of the perturbative expansion in the  $\overline{\hbox{MS}}$ scheme. 
 
 In the present paper, we  extend  the previous analysis to  the series obtained in the zero momentum massive scheme in $d=3$ at six loops \cite{pelissetto01a}  and in $d=2$  at five loops \cite{calabrese03}.  In $d=3$, we apply the criteria   used in  \cite{mudrov98c,delamotte08}   --- Principle of Minimal Sensitivity (PMS) and Principle of Fastest Apparent Convergence (PFAC) ---   and confirm that the fixed points  found in $d=3$   for $N=2$ and $N=3$ are most likely spurious. In $d=2$, the situation is more delicate.  We  recall that,  already for the  (non-frustrated)  $O(N)$ models, the critical exponents found  from the $(\pmb \phi^2)^2$ field theory  are quantitatively wrong although apparently converged for all $N$. This  striking   phenomenon, already  mentioned   in \cite{orlov00}, relies on  the presence of  non-analytic contributions  to  the  $\beta$-function at the fixed  point \cite{calabrese00}.   The same  kind of problem has  been  mentioned  in the case of frustrated magnets \cite{calabrese01,calabrese03}  but  it  was assumed  to  leave  unaffected  the qualitative  predictions, in particular,   the existence of a non-trivial fixed point in the Heisenberg case.      We  show here, on the contrary,  that the phenomenon of apparent convergence towards erroneous  values, together with the presence of instabilities of  some critical exponents with respect to the resummation parameters,  leads   to   seriously  question   the conclusions drawn  in the past  as for the critical behaviour of these systems.

 Our study  altogether  shows   clear evidences  that  the FD perturbative approaches  to   three-dimensional  frustrated magnets  are not reliable, at least at the orders studied,  and that there is no convincing  evidence  of a   genuine   phase transition induced by vortices  in   two-dimensional Heisenberg spin systems.  Finally, for the same reasons as in the Heisenberg case, we show that  the behaviour  of XY spins  in $d=2$  cannot be elucidated from the  five-loop perturbative data.

The paper is organized as follows. In  Section II, we study  in detail  the $O(N)$ case in $d=2$ and $d=3$.  In $d=3$, this  allows us to illustrate  on a well-known example the kind  of stability (resp. instability) properties expected for a  genuine  (resp. spurious) fixed point. 
In $d=2$, this allows us to  illustrate  the fact that there can be  fast apparent convergence of the critical exponents  but towards  erroneous  values  due to nonanalytic contributions. 
In  Section III,  frustrated magnets are  studied in $d=3$ and in $d=2$. In $d=3$,  we confirm the spurious character of the fixed points found for $N=2$ and $N=3$. In $d=2$,   we  show the unreliability of the conclusions   -- phase transition controlled by a fixed point -- deduced from the results  obtained perturbatively at five loops in the $N=3$ case.  We then  analyze  the $N=2$ case and reach the same conclusions as in the $N=3$ case.

\section{The $O(N)$ models  in two and three dimensions}

In the following, we study the convergence of the resummed perturbative series obtained for the frustrated models.
Since  we  need to determine  criteria to decide whether  the  perturbative results are (or are not) converged,   we illustrate briefly how convergence of the resummed series shows up for  the $O(N)$ models  in $d=3$.   We show that the  behavior  of   the correction to scaling exponent $\omega$ 
as a function either of the loop order or of the resummation parameters  is a good indicator of the numerical convergence of the perturbative results. 
The exponent $\eta$,  when available,  is also studied. By analyzing  the two-dimensional $O(N)$ case we  also show that, contrarily to common belief,  the five-loop results for the critical exponents are not converged.   The reason of this behaviour  is however rather subtle since
there is, in fact,  apparent convergence but towards erroneous values, a phenomenon that we call  {\sl  anomalous apparent  convergence}.

\subsection{Resummation procedure}

As well known, the perturbative series obtained in the $O(N)$ models  for the $\beta$ function describing 
the running of the coupling constant with the scale  are not convergent \cite{hardy48,zinnjustin89}.   They are  asymptotic series which, 
in the case of the zero momentum massive scheme,  are  Borel summable  \cite{magnen77}. Powerful resummation methods have been used in the past that, thanks to the knowledge of the behavior of the series at large order, lead to converged and accurate results (see \cite{zinnjustin89, suslov05} for  reviews).  We recall in the following the kind of resummation procedure that we use throughout this article.

 Let  us consider  a
 series
\begin{equation}
f(u)=\sum_{n} a_n \ u^n \
\label{series1}
\end{equation}
where the coefficients $a_n$ are supposed to grow as $n!$.

The Borel-Leroy sum associated with $f(u)$ is given by:
\begin{equation}
B(u)=\sum_{n} {a_n\over \Gamma[n+b+1]} \ u^n \ 
\label{borelsum}
\end{equation}
 where $b$ is a  parameter whose meaning will become clear later.

The resulting series  is now supposed to converge, in the complex  plane, inside a circle of
radius $1/a$, where  $u=-1/a$ is the  singularity of $B(u)$ closest to
the  origin.   Then,        using  this   definition  as    well    as
$\Gamma[n+b+1]=\int_0^{\infty} t^{n+b}\ e^{-t} dt$, one can rewrite
\begin{equation}
f(u)= \sum_{n} {a_n\over \Gamma[n+b+1]}  \ u^n \int_0^{\infty} \  dt \
e^{-t}\ t^{n+b}\ .
\end{equation}
Interchanging summation and integration, one can now {\it  define} the
Borel transform of $f$ as:
\begin{equation}
f_B(u)=\int_0^{\infty} \ dt \ e^{-t}\ t^{b}\ \ B(ut)\ .  \
\label{boreltrans}
\end{equation}

In order  to perform the integral (\ref{boreltrans})  on the whole
real positive  semi-axis one has to  find  an analytic continuation of
$B(u)$.    Several methods  can   be   used, Pad\'e   approximants  constitute one possibility  \cite{baker78,baker96,holovatch02}.  
 However,   it is generally   believed  that the   use of a
conformal  mapping \cite{leguillou77,leguillou80} is more  efficient   since  it makes  use  of  the
convergence  properties of the Borel  sum.   Under the assumption that
all the singularities of $B(u)$ lie on the negative real axis and that
the Borel-Leroy sum is analytic in  the whole complex plane except for
the cut extending from $-1/a$ to $-\infty$, one can perform the change
of variable:
\begin{equation}
\omega(u)={\sqrt{1 + a\, u}-1\over \sqrt{1 + a\, u}+1}
 \hspace{1cm}  \Longleftrightarrow     \hspace{1cm}  u(\omega)={4\over
 a}{\omega\over(1-\omega)^2}
\label{conformal}
\end{equation}
that maps the  complex $u$-plane cut from  $u=-1/a$  to $-\infty$ onto
the unit circle in the $w$-plane such that the singularities of $B(u)$
lying  on the negative axis   now lie on   the boundary of the  circle
$|w|=1$.  The   resulting expression $B(u(\omega))$ has   a convergent
Taylor  expansion  within the  unit circle   $|\omega|<1$  and can  be
rewritten:
\begin{equation}
B(u(\omega))=\sum_{n} d_n(a,b) \hspace{0.1cm} \left[\omega(u)\right]^n
\label{borel3}
\end{equation}
where    the coefficients $d_n(a,b)$  are      computed so  that   the
re-expansion  of   the r.h.s.  of    (\ref{borel3})  in powers of  $u$
coincides  with  that   of   (\ref{series1}).   One  obtains   through
(\ref{borel3}) an analytic  continuation  of $B(u)$ in  the  whole $u$
cut-plane  so that  a resummed  expression of  the  series  $f$ can be
written:
\begin{equation}
f_R(u)=\sum_{n} d_n(a, b) \hspace{-0.1cm} \int_0^{\infty}
\hspace{-0.2cm}dt\, \, {e^{-t}\, t^{b}\ \left[\omega(u t)\right]^n}\ . 
\label{resummation1}
\end{equation}
        
In practice it  is   interesting  to   generalize  the
expression  (\ref{resummation1}) by   introducing \cite{kazakov79} the
expression

\begin{equation}
f_R(u)=\sum_{n} d_n(\alpha,a, b) \hspace{-0.1cm} \int_0^{\infty}
\hspace{-0.2cm}dt\, \, {e^{-t}\,  t^{b}}\ { \left[\omega(u t)\right]^n \over 
\left[1-\omega(u t)\right]^{\alpha} }\  
\label{resummation2}
\end{equation}
whose meaning  will be explained  just below.

If  an infinite  number of terms   of the series  $f_R(u)$ were known,
expression (\ref{resummation2}) would be independent of the parameters
$a$,  $b$ and $\alpha$. However  when only a  finite number of terms
are known, $f_R(u)$  acquires a dependence  on them. In principle, the
parameters $a$  and $b$ are  fixed by the  large order behavior of the
series:
\begin{equation}
a_{n\to\infty}\sim (-a_{\rm lo})^n \, n!\, n^{b_{\rm lo}}
\label{a-large-order}
\end{equation}
which  leads  to $a=a_{\rm lo}$   and  $b\gtrsim b_{\rm lo}+3/2$ \cite{leguillou80} where 
 $a_{\rm lo}$  and $b_{\rm lo}$   denote  the large-order value of $a$ and $b$. As for
$\alpha$, it is determined by the  strong coupling behavior of the initial
series:
\begin{equation}
f(u\to\infty) \sim u^{\alpha_0/2} \
\end{equation}
which can  be  imposed  at any  order of   the  expansion  by choosing
$\alpha=\alpha_0$.  The common assumption is that  the above choice of
$a$, $b$ and  $\alpha$   improves the convergence  of   the resummation
procedure since it encodes exact results.

 Let us however emphasize that, often, only  $a$ is known and that the
 other parameters, $\alpha$ and $b$, must be considered either as free (as
 for instance in  \cite{calabrese04}) or variational (as for  instance in
 \cite{mudrov98c,delamotte08} where  $\alpha$   is determined by    optimizing the
 apparent convergence of the series). In  any case,  the choice of value
 of  $a$, $\alpha$ and $b$ must  be validated a posteriori.

\subsection{$O(N)$  models in three dimensions and  principles of convergence}

The  dependence  of the critical exponents upon the parameters  $a, b$ and $\alpha$ is an indicator of   the (non-) convergence of the perturbative series. Indeed, in  principle,  any 
converged physical quantity $Q$ should be independent of these parameters.  However, in practice, at a given order $L$ of approximation
(loop order), all physical quantities depend (artificially) on them: $Q\to Q^{(L)}(a,b, \alpha)$. 
Even if $a$ is fixed at the value obtained from the large order behavior,
all physical quantities remain dependent  upon  $ b$ and $ \alpha$ at finite order. 
We consider that the optimal result for $Q$
at order $L$  corresponds to the values $(b_{\rm opt}^{(L)},\alpha_{\rm opt}^{(L)})$ of $(b, \alpha)$  for which Q  depends most weakly  on $b$ and $\alpha$, {\it  i.e.} for which   it  is stationary:
\be
Q^{(L)}_{\rm opt}= Q^{(L)}(b_{\rm opt}^{(L)},\alpha_{\rm opt}^{(L)})\ \ \ \ \ \ {\rm with}\ \ \ \ \ \ 
\frac{\partial Q^{(L)}(b,\alpha)}{\partial b}\,{\bigg\vert_{b_{\rm opt}^{(L)},\alpha_{\rm opt}^{(L)}}}=\frac{\partial Q^{(L)}(b,\alpha)}{\partial\alpha}\,{\bigg\vert_{b_{\rm opt}^{(L)},\alpha_{\rm opt}^{(L)}}}=0\ 
\label{PMS}
\ee
where, of course, $b_{\rm opt}^{(L)}$ and $\alpha_{\rm opt}^{(L)}$ are functions of the order $L$.

The validity of this procedure, known as the ``Principle of Minimal Sensitivity'' (PMS), requires 
that there is a unique pair  $(b_{\rm opt}^{(L)},\alpha_{\rm opt}^{(L)})$ such that $Q^{(L)}$ is stationary. This is generically not  the case:  several stationary points are often found. A second principle allows us to ``optimize'' the results even in the case where there are several ``optimal'' values of $b$ and $\alpha$ at a given order $L$:
this is the so-called ``Principle of Fastest Apparent Convergence'' (PFAC). 
The idea underlying this principle is that
when the numerical value of $Q^{(L)}$ is almost converged (that is $L$ is sufficiently large to achieve 
a prescribed accuracy) 
then the next order of approximation must consist only in  small change
of this value: $Q^{(L+1)}\simeq Q^{(L)}$. Thus, the preferred values of $b$ and $\alpha$ should be  the ones
for which  the difference between two successive orders  $Q^{(L+1)}(b^{(L+1)},\alpha^{(L+1)})- Q^{(L)}(b^{(L)},\alpha^{(L)})$ is minimal. 
In practice, the two  principles should be used together  for consistency and,  if there are several
solutions to Eq.(\ref{PMS}) at order $L$ and/or $L+1$, one should  choose the couples   $(b_{\rm opt}^{(L)},\alpha_{\rm opt}^{(L)})$  and $(b_{\rm opt}^{(L+1)},\alpha_{\rm opt}^{(L+1)})$ for which the stationary values  $Q^{(L)}(b_{\rm opt}^{(L)},\alpha_{\rm opt}^{(L)})$ and 
$Q^{(L+1)}(b_{\rm opt}^{(L+1)},\alpha_{\rm opt}^{(L+1)})$  are the closest, that is for which there  is fastest apparent convergence.   These principles have  been developed and used in \cite{leguillou80,mudrov98c,delamotte08}, see also \cite{dudka04}.

\begin{figure}[htbp]
\begin{center}
\includegraphics[width=2.2in,origin=tl]{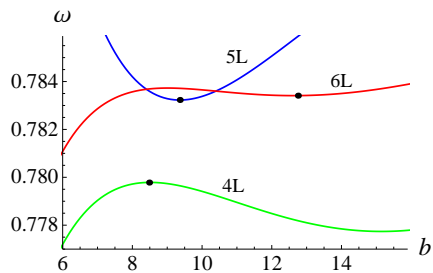}\hfill%
\end{center}
\caption{The exponent $\omega$ of the three-dimensional $O(4)$ model as a function of the resummation parameter $b$ 
at  four-, five- and six-loop orders.  The dot  on each curve corresponds to a stationary value of $\omega=\omega(\alpha,b)$ in both  $\alpha$ and $b$ directions  with $a$ fixed to its large-order value $a_{\rm lo}\simeq 0.1108$. One has: $(\alpha_{\rm opt},b_{\rm opt})=(3.2,8.5)$  at four loops,   $(\alpha_{\rm opt},b_{\rm opt})=(5.2,9.5)$  at five  loops and  $(\alpha_{\rm opt},b_{\rm opt})=(5,13)$  at six loops.}
\label{omegaO43db}
\end{figure}

Nice examples where these two principles work very well and indeed lead to optimized values
of the critical exponents are the $O(N)$ models  in $d=3$ computed perturbatively at  four-,
 five- and six-loop orders (within the zero momentum massive scheme).
 The series for the $\beta$-function
of the coupling constant are resummed thanks to a conformal Borel transform.  Subsequently, one obtains the fixed point coordinate $u^*$,
its stability being defined by the correction to scaling exponent $\omega$:
\begin{equation}
\omega=\left.\frac{\partial \beta(u)}{\partial u}\right|_{u=u^*}.
\end{equation}
A positive value of $\omega$ (or  a positive real part if it is complex) corresponds to a stable fixed point. We show in  Fig.\ref{omegaO43db}  the exponent $\omega$  of the $O(4)$ model in $d=3$ as a function of the parameter $b$  for the values of $\alpha$  for which  stationarity  is found for both $b$ and $\alpha$.  
As expected, the dependence of $\omega$  upon the resummation parameters becomes 
smaller as the order in the loop expansion increases  as illustrated  by  the curves $\omega(b)$ that  flatten  between four  and  six loops, see Fig.\ref{omegaO43db}.  At this order one  finds $\omega\simeq  0.783$.   As an indicator of the quality of the convergence we  give the   difference between the fifth and the sixth order  for the exponent $\omega$:  $\omega(L=6)-\omega(L=5)\simeq  2\ .\  10^{-4}$.  Note that  this case also illustrates the  situation where -- at six-loop order -- several stationary points  occur and where  the PFAC allows us to select a single solution, see Fig.\ref{omegaO43db}.  Our results are  comparable with six-loop results obtained  by Guida and Zinn-Justin \cite{guida98} for $N=4$: $\omega=0.774\pm0.020$ (in $d=3$),  $\omega=0.795\pm0.030$ (within the $\epsilon$-expansion). 

It is remarkable that the same study  performed on other critical exponents, other values of 
$N$ and even with other perturbative series (obtained from the $\overline{\hbox{MS}}$ scheme for instance)  always leads to the same kind of results   with  values of the exponents   found    that are very close  to the best known values obtained
from Monte Carlo simulations. This proves that the above methodology is indeed  efficient.

\begin{figure}[htbp]
\begin{center}
\includegraphics[width=2.2in,origin=tl]{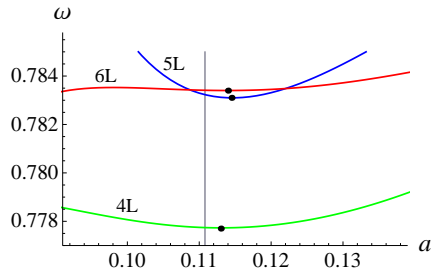}\hfill%
\end{center}
\caption{The exponent $\omega$  of the three-dimensional $O(4)$ model  as a function of the resummation parameter $a$ at  four, five  and six loops. The vertical line corresponds to  $a=a_{\rm lo}\simeq 0.1108$. 
The values chosen  for $\alpha$ and $b$ are such that $\omega$ is stationary   w.r.t. $\alpha$ and $b$ when $a=a_{\rm lo}$  (see Fig.\ref{omegaO43db}). Note  that $\omega(a_{\rm opt})\simeq \omega(a_{\rm lo}$).} 
\label{omegaO43da}
\end{figure}

 Finally note that the  same argument can also be applied  to the determination of  an optimal value of $a$, $a_{\rm opt}$,  from the PMS applied to this parameter:
if there is convergence {of the resummed series}, we expect that  $a_{\rm opt}$ almost coincides with the value determined
 by the large order analysis, Eq.(\ref{a-large-order}), $a_{\rm opt}\simeq a_{\rm lo}$. The difference between these quantities  is a measure of the convergence level of the series. We show in Fig.\ref{omegaO43da} on the example of the $O(4)$ model in $d=3$
that, as expected, the value  $a_{\rm opt}$  is very close to $a_{\rm lo}$ and the difference between $\omega(a_{\rm opt})$ and $\omega(a_{\rm lo})$ 
is extremely small.

\begin{figure}[htbp]
\begin{center}
\begin{picture}(500,110)
\put(80,-9) {(a)}
\put(10,5){\includegraphics[width=0.37\textwidth]{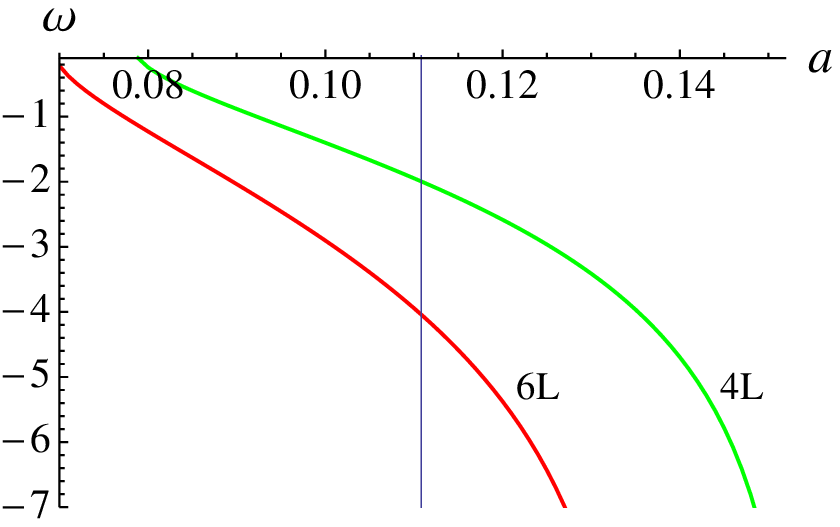}}
\put(320,-9){(b)}
\put(245,-4){\includegraphics[width=0.37\textwidth]{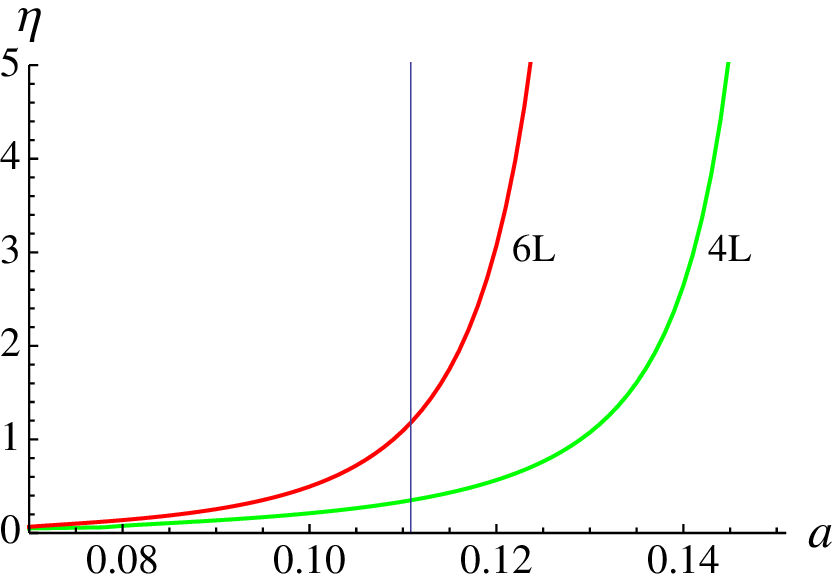}}
\end{picture}
\end{center}
\caption{The exponents  $\omega$ and $\eta$ as functions of $a$ at the spurious fixed point of the $O(4)$ model in $d=3$ at  four   and  six loops. The vertical line corresponds to  $a=a_{\rm lo}\simeq 0.1108$.  The values considered  for $\alpha$ and $b$ are respectively equal to $6$ and  $4$.  Other values gives similar results. The exponent $\omega$ is negative since the
spurious fixed point is repulsive.}
\label{spure_a}
\end{figure}

 The criteria  given above are of crucial importance, especially  when considering FD approaches.  Indeed, generically the (non-resummed) series obtained  at $L$ loops for the $\beta$-function are polynomials of order $L+1$ in the coupling constant $u$. Thus, the fixed point equation $\beta(u^*)=0$  admits  $L+1$ roots $u^*$ that are either real or complex.   Contrary to what occurs within the $\epsilon$-expansion,  where the  coupling constant is by definition of order $\epsilon$,  in the FD approach  no   real  root can be  {\sl a priori} selected or, reciprocally, discarded. As a result, the generic  situation is that  the number of  fixed points as well as their stability vary  with the order $L$:  at a given  order, there can exist several real and stable fixed points or none instead of a single one. In principle,  the resummation procedure  allows both to restore the non-trivial  Wilson-Fischer fixed point and to eradicate the non-physical, spurious,  roots. In particular,  we expect  that spurious solutions should   satisfy neither the PMS nor the PFAC criteria.  We show in Fig.\ref{spure_a} on the example of the 3$d$ $O(4)$ model  that  there exists,  beside the usual Wilson-Fisher stable fixed point,  a  spurious  (unstable) fixed point. As expected,  the exponents computed  at this fixed point  are very unstable w.r.t. variations of the resummation parameters $a$ (the same behaviour occurs when variations of  $b$ are considered),  a behaviour  which,  according to our criteria,  is sufficient to discard it \footnote{Notice that the existence of a spurious fixed point depends on the parity of $L$.}.

  A potential   difficulty  with  this procedure  is  that  for more involved  models, bringing into play several  coupling constants,   the resummation  procedure is very likely less efficient than for the $O(N)$ models since it is   performed with respect to one coupling constant only -- see below.  In this case,  the  instability  displayed by  the spurious fixed points can be  much  weaker   than for  $O(N)$ models.   However some of the  present authors have previously shown \cite{delamotte08} that the  criterion of (in)stability given above still  remains reliable in these  more general and ambiguous situations.  More precisely they have shown, on the example of  frustrated magnets (and on the model with cubic anisotropy)     that    fixed points   suspected  to  be  spurious  from a  stability analysis,  have been   confirmed to be so  from additional  independent  arguments. These arguments are:   (i) persistence of the fixed point as a non-trivial,  non-Gaussian  one up to (and above)  the upper critical   dimension $d=4$,  a  fact which is  forbidden for a $\phi^4$-like theory  (see \cite{kenna93,suslov08}  and reference therein),   (ii)   existence of a topological singularity in the mapping between $(N,d)$  and the  fixed point coordinates that makes these  last quantities multivalued  functions of $(d,N)$  that is manifestly  a pathology.

From the discussion above it appears  that a necessary condition for a fixed point  to be considered as a genuine fixed point
is that it satisfies both  the PMS and  the PFAC. We however now show
on the example of the 2$d$ $O(N)$ models  that,  although necessary, these conditions are    not sufficient.

\subsection{$O(N)$  models  in two dimensions:  anomalous apparent convergence}

\begin{figure}[ht]
\begin{center}
\begin{picture}(500,110)
\put(100,-5) {(a)}
\put(10,5){\includegraphics [width=0.37\textwidth]{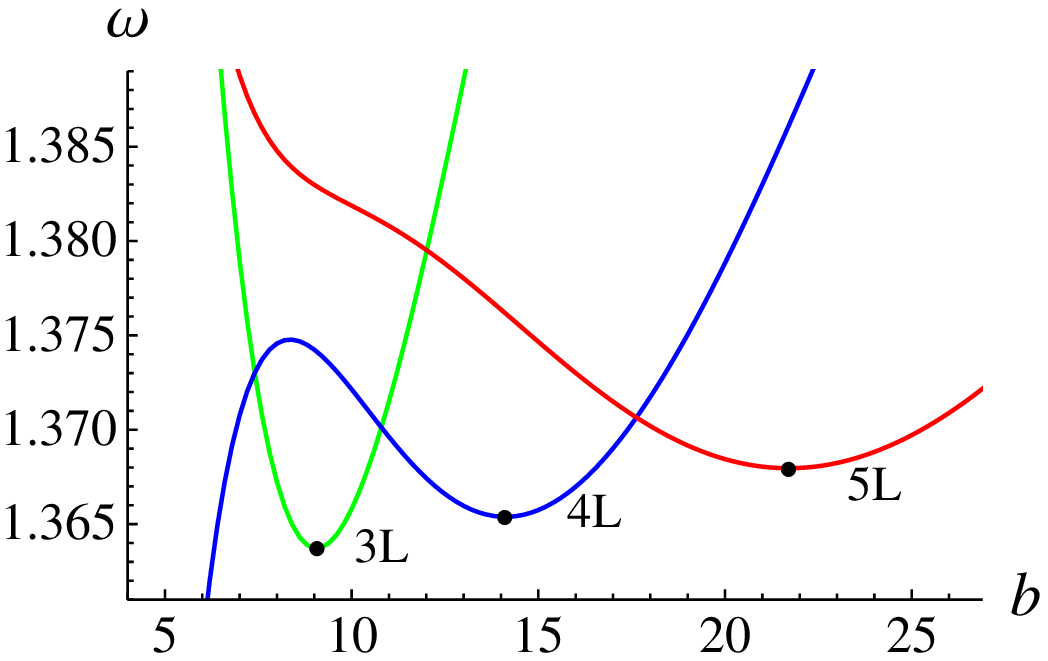}}
\put(330,-5){(b)}
\put(245,5){\includegraphics[width=0.37\textwidth]{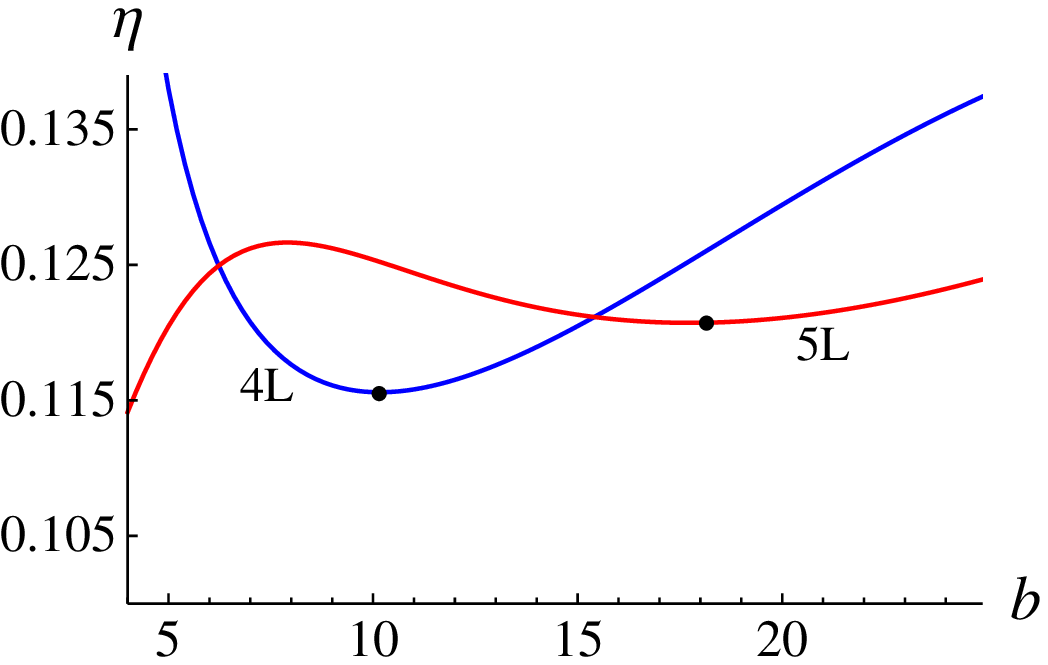}}
\end{picture}
\end{center}
\caption{The exponents $\omega$ and $\eta$ of the two-dimensional $O(4)$ model as 
functions of the resummation parameter $b$ at three-, four- and five-loop order  (the result for $\eta$ at three loops is not displayed since there is no clear stationarity for $\eta$ in this case). The parameter $a$ has been fixed at the value obtained 
from the large order behavior: $a=a_{\rm lo}\simeq 0.1789$. For $\omega$ one has:  $(\alpha_{\rm opt},b_{\rm opt})=(3.1,9)$ at three  loops, $(\alpha_{\rm opt},b_{\rm opt})=(3.1,14)$  at four  loops and $(\alpha_{\rm opt},b_{\rm opt})=(3.1,21.5)$  at five  loops. For $\eta$ one has:  $(\alpha_{\rm opt},b_{\rm opt})=(4.4,10)$ at four loops and $(\alpha_{\rm opt},b_{\rm opt})=(4.6,18)$  at  five loops.   The dot on each curve  corresponds  to  stationary values of $\omega=\omega(\alpha,b)$ and $\eta=\omega(\alpha,b)$
in both  directions.}
\label{omega_eta_2d}
\end{figure}

The same kind of analysis of the perturbative results obtained from the $\phi^4$ model 
 in three dimensions   can be performed for all $N$ in two dimensions.  We show in  Fig.\ref{omega_eta_2d} the exponent $\omega$  of the two-dimensional $O(4)$ model obtained  at three, four and five loops in the zero-momentum massive scheme and the  anomalous dimension  $\eta$ at  four and five loops (the three loops results does not lead to a clear stationary behaviour).  A conformal Borel resummation method has been used. 

For this model,  because  of  Mermin-Wagner's theorem \cite{mermin66}, the correlation length is  infinite at  zero temperature only
and the critical exponents are exactly known: $\eta=0$ and $\omega=2$  \cite{polyakov75,brezin76}. We can see on Fig.\ref{omega_eta_2d}
that although the values obtained for these  exponents seem  well converged, they are    erroneous    since  using both
the PMS and PFAC one finds: $\eta\simeq 0.12$
and $\omega\simeq 1.37$. It is important
to emphasize that $N=4$ is not an isolated case in this respect. For all two-dimensional  $O(N)$ -- with $N\ge 1$  -- models the critical exponents seem
to be converged at five-loop order but towards  erroneous  values.
For instance, in the Ising model,  and at five loops, Orlov and Sokolov \cite{orlov00} have  found    $\eta=0.146$,  and Pogorelov and Suslov \cite{pogorelov07}  $\eta=0.145 (14)$  whereas the exact result is $\eta=0.25$.

\begin{figure}[h]
\begin{center}
\begin{picture}(500,110)
\put(90,-5) {(a)}
\put(10,5){\includegraphics[width=0.37\textwidth] {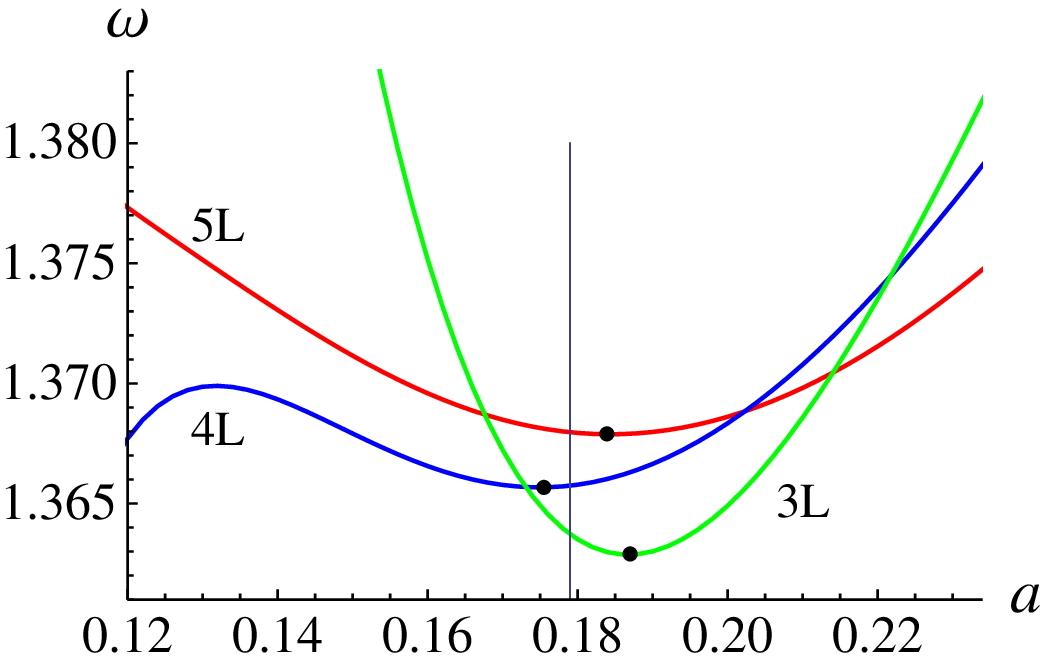}}
\put(320,-5){(b)}
\put(245,5){\includegraphics[width=0.37\textwidth] {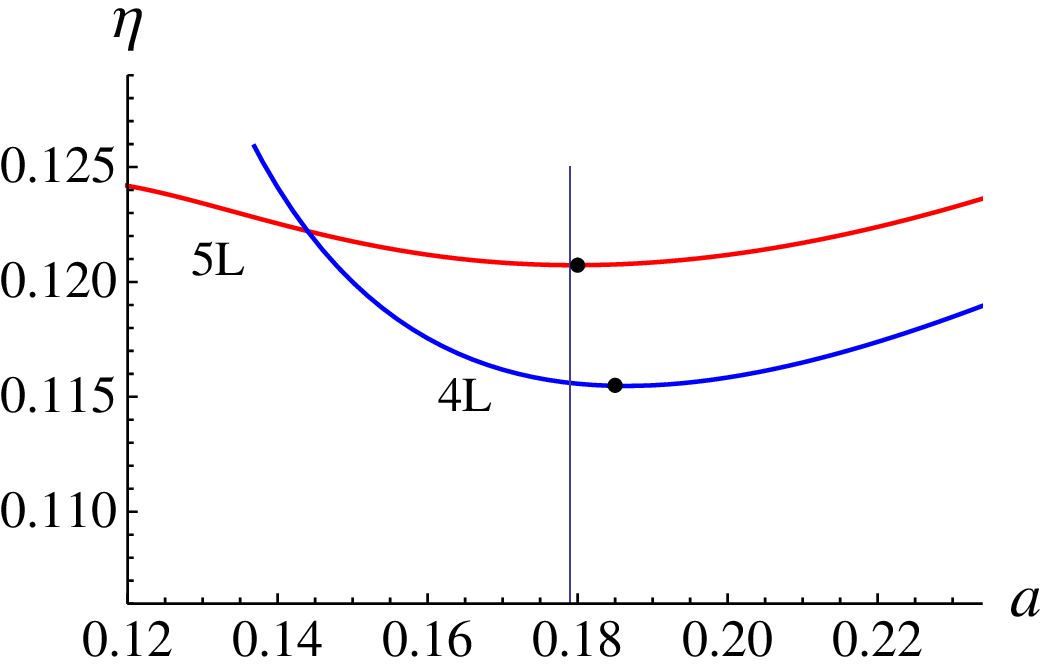}}
\end{picture}
\end{center}
\caption{The exponents  $\omega$ and $\eta$ of the two-dimensional $O(4)$ model as  functions  of the resummation parameter $a$, at  three, four and five loops (the result for $\eta$ at three loops is not displayed since there is no clear stationarity  in this case). The vertical line corresponds to  $a_{\rm lo}\simeq 0.1789$.
The values chosen for $\alpha$  and $b$  are such that the exponents are at their stationary value when $a = a_{\rm lo}$ (see Fig.\ref{omega_eta_2d}).   Note that the stationary values $\omega(a)$ and $\eta(a)$  indicated by dots are very close to  their values at $a=a_{\rm lo}$. }
\label{omegaO4a}
\end{figure}

We have  also studied the $a$-dependence  of  the critical exponents.  We have found here again that $a_{\rm opt}\simeq a_{\rm lo}=0.1789$, 
see Fig.\ref{omegaO4a}. This means that  the $a$-dependence is not either a good indicator of an anomalous apparent convergence in this case.


An analysis of the underlying reasons of this anomalous  convergence has been performed  in  the Ising case in \cite{calabrese00} 
(see also \cite{sokal94,sokal95}).  The explanation is that there very likely  exist,  in  the $\beta$-function,  terms
such as $1-(1-u/u^*)^e$ with  $u^*$ the fixed point value of $u$  and $e$ a small number (probably 1/7 in the Ising case) \cite{calabrese00,pelissetto01c}. In the perturbative expansion performed around $u=0$, such terms lead to  small  contributions 
to the $\beta$-function  that  seems  to be  under control.   However  they play an important role in 
the vicinity of $u^*$;  they are even non-analytic at this point since their derivatives  with respect to $u$  are singular at  $u^*$. 
Reconstructing such terms from a perturbative  expansion is thus difficult and, as a consequence, the perturbative results 
are doomed to failure although they look converged.   Thus we are lead to the conclusion  that PMS and PFAC are necessary 
conditions for convergence but are not sufficient.

\begin{figure}[htbp]
\begin{center}
\begin{picture}(500,110)
\put(80,-5) {(a)}
\put(10,5){\includegraphics[width=0.37\textwidth]{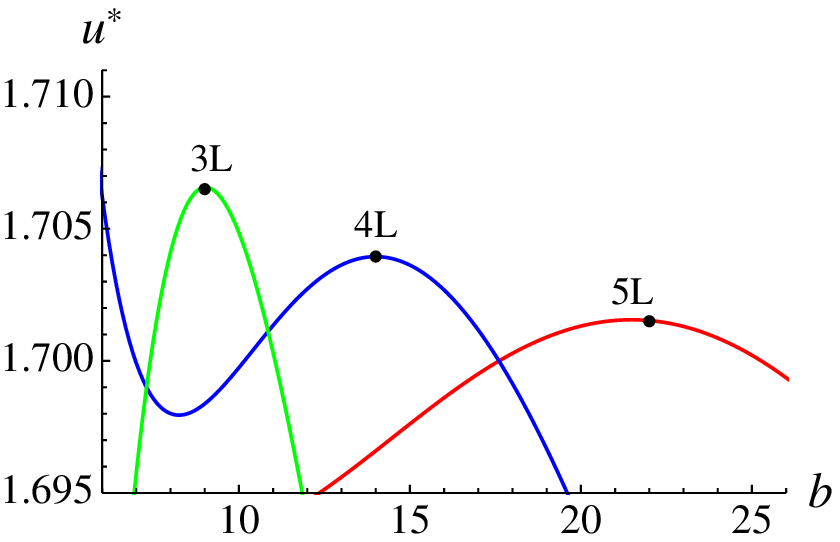}}
\put(320,-5){(b)}
\put(245,5){\includegraphics[width=0.37\textwidth]{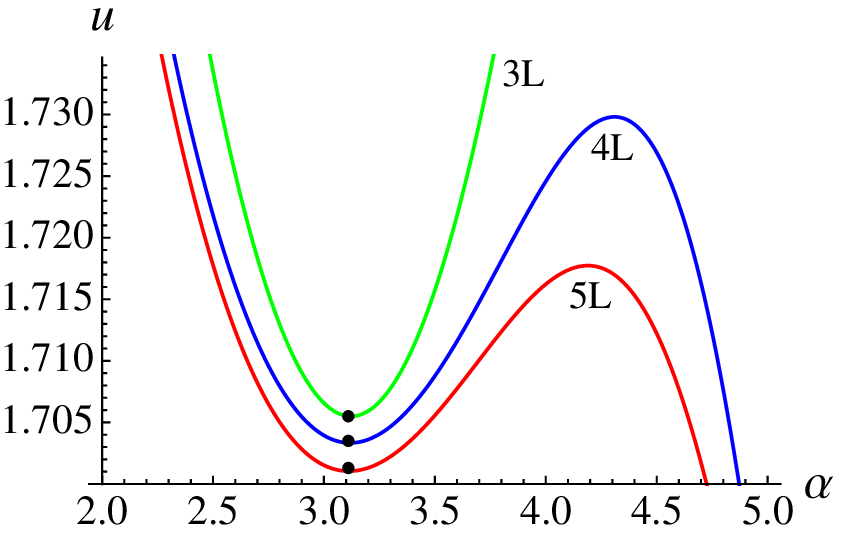}}
\end{picture}
\end{center}
\caption{The fixed point coupling constant  $u^*$  of the  $O(4)$ model in $d=2$: (a) as a function of $b$ at fixed $\alpha=3.1$,  
 (b) as a function of $\alpha$ at fixed $b=b_{\rm opt}=9, 14, 22$ at three, four and five  loops respectively.
 The parameter $a$ has been fixed at $a_{\rm lo}=0.1789$.}
\label{uO4alpha}
\end{figure}

Let us now  make a remark  specific to  $d=2$. In this dimension,  the existence  of a non trivial  root $u^*$  of  the $\beta$ function, stable with respect to the resummation parameters $b$ and $\alpha$ and  displaying good convergence properties,  see Fig.\ref{uO4alpha},
is not  sufficient in  itself to  know   whether the transition  is trivial (taking  place  at  zero temperature) or not,  since $u$ is not directly related to the temperature.  In principle, the triviality (for $N\ge3$)  or non triviality (for $N=1$ or 2) of the critical exponents  should be sufficient  to conclude.  However, as previously emphasized, the presence of strong non-analyticities in the two-dimensional $\beta$-functions of the Ising and $O(N)$ models
prevent us to do so since  they completely spoil the determination of the critical exponents.

\section{The frustrated $O(N)\times O(2)$ models}

Let us now come to the frustrated models we are directly interested in.  A first analysis of the convergence of the ${\overline{\hbox{MS}}}$ series obtained  at five loops in these models was done in  \cite{delamotte08}. In the following, we  study 
these models by analyzing the perturbative series obtained in the massive zero momentum scheme at six loops in three dimensions and at five loops in two dimensions,  a case   that   the $\overline{\hbox{MS}}$ series do  not  allow   to satisfactorily  study  since  the   values of the coupling constants  at the fixed point are  out of the region  of Borel-summability.

  The Hamiltonian  relevant for frustrated
 systems is given by \cite{garel76,bailin77,yosefin85,kawamura88}:
\begin{equation}
\begin{array}{ll}
\displaystyle \hspace{0cm}{\mathcal H}= \int{\rm d^d} x \Big\{\frac{1}{2}
\left[(\partial\pmb \phi_1)^2+
 (\partial\pmb\phi_2)^2 + m^2 (\pmb\phi_1^2+\pmb\phi_2^2)\right]+\\
\\
\hspace{1cm}\displaystyle \frac{u}{4!}[\pmb\phi_1^2+\pmb\phi_2^2]^2 
+\frac{v}{12}[(\pmb\phi_1 \cdot \pmb\phi_2)^2-
\pmb\phi_1^2\pmb\phi_2^2 ] \Big \}
\end{array}
\label{landau}
\end{equation}
where $\pmb\phi_i$, $i=1,2$ are $N$-component vector fields. The resummation procedure 
outlined above can be generalized to the case where there are several coupling constants
as it is the case for frustrated systems.
For a function $f$  of the two
variables  $u$ and $v$ known through its  series  expansion in
powers   of $u$  and $v$, the resummation procedure used in \cite{pelissetto01a,carmona00,calabrese01}
consists in  assuming that $f$ can be considered as a
function of $u$ and of the ratio $z=v/u$:
\begin{equation}
f(u,z)=\sum_{n} a_n(z) \ u^n \ \label{series}
\end{equation}
and in resumming with respect to  $u$  only.
Under this hypothesis  the   resummed expression associated with
$f$ reads:
\begin{equation}
f_R(u,z)=\sum_{n} d_n(\alpha,a(z),b;z) \hspace{-0.1cm}
\int_0^{\infty} \hspace{-0.2cm}dt\, \, {e^{-t}\, t^{b}}{
\left[\omega(u t;z)\right]^n \over \left[1-\omega(u
t;z)\right]^{\alpha} } \label{resummation}
\end{equation}
with:
\begin{equation} \label{om}
\omega(u;z)={\sqrt{1 + a(z)\, u}-1\over \sqrt{1 + a(z)\, u}+1}\
\end{equation}
where,  as     above,  the  coefficients
$d_n(\alpha,a(z),b,z)$ in (\ref{resummation})  are computed  so
that   the  re-expansion of  the r.h.s.  of (\ref{resummation}) in
powers of $u$ coincides with that of (\ref{series}). Of course, since the resummation is
performed in only one variable, we cannot expect in this case a convergence of the resummed
quantities as good as in the $O(N)$ case.

\subsection{The frustrated models in $d=3$}

\begin{figure}[htbp] 
\vspace{0cm}
\hspace{2cm}
\includegraphics[width=0.5\linewidth,origin=tl]{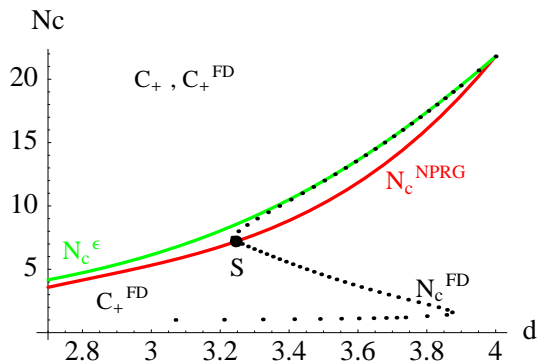}\hfill%
\caption{Curves $N_c(d)$ obtained within the $\epsilon$-expansion ($N_c^\epsilon$), the  $\overline{\hbox{MS}}$
 scheme  without $\epsilon$-expansion ($N_c^{\text{FD}}$) and the NPRG approach ($N_c^{\text{NPRG}}$). 
The resummation parameters for the $\overline{\hbox{MS}}$ curve are $a=0.5$, $b=10$ 
and $\alpha=1$. The  part of the  curve $N_c^{\text{FD}}$ below S  corresponds
 to a regime of non-Borel-summability. The attractive fixed point is called $C_+$ and ${C_+}^{\rm FD}$ when it is found
in the fixed dimension  ($\overline{\hbox{MS}}$) scheme.  } 
\label{courbes_ncd} 
\end{figure}

We recall in Fig.\ref{courbes_ncd} the results obtained 
using  different approaches. In the $(d,N)$ plane a line  $N_c(d)$ is found in all approaches
such that    the stable fixed point exists for $N>N_c(d)$ and disappears  for $N<N_c(d)$. This result is interpreted as the occurrence of  a second order  transition for  values of $N$ above $N_c(d)$ and a first order transition for values of $N$  below  $N_c(d)$. In the $\epsilon$-expansion \cite{antonenko95,holovatch04,calabrese03c} and within the NPRG  \cite{tissier00,tissier00b,tissier01, tissier01c, delamotte03} the lines   $N_c(d)$ are both monotonic and are very similar  (see  Fig.\ref{courbes_ncd}).  They lead to the fact  that $N_c(d=3)>3$  and the transition is thus found to be of first order for $N=2$ and 3 in three dimensions.
On the contrary, in the $\overline{\hbox{MS}}$ scheme without $\epsilon$-expansion \cite{calabrese04}, the curve  $N_c(d)$ is found to have a $S$-shape, see Fig.\ref{courbes_ncd}, and thus,  at $d=3$,  fixed points exist  for $N=2$ and 3. In the massive scheme  also
fixed points are found for these values of $N$ \cite{pelissetto01a}.

The $\overline{\hbox{MS}}$ scheme perturbative series at five loops were reexamined in  \cite{delamotte08}. 
The bad convergence of the resummed series, the analytic properties of the coordinates of the 
fixed points $(u^*,v^*)$ considered as functions of $(d,N)$  (presence of a topological
singularity $S$, see Fig.\ref{courbes_ncd}, in the $(d,N)$ plane) and the fact that
the fixed points found at $N=2$ and $N=3$ in $d=3$ do not become Gaussian when they are followed
continuously in $d$ up to $d=4$ led the authors of  \cite{delamotte08} to conclude 
that these fixed points were either spurious or the results non converged. By re-analyzing the resummed series obtained at six loops 
in $d=3$ in the massive scheme we  show in the following (i) that for sufficiently large values of $N$ (typically $N>7$)
the resummed series for the exponents converge well, (ii) that for $N=2$ and 3 these series  do not 
lead to converged results. The situation is thus similar to what has already been obtained in the 
$\overline{\hbox{MS}}$ scheme.

\subsubsection{The $N=8$  frustrated model in $d=3$}

Let us start our analysis by the $N=8$ case to show how   the results obtained
at large and small values  of $N$ for frustrated systems display completely  different  convergence properties. Since the model involves two coupling constants $u$ and $v$ 
there are two eigenvalues of the stability matrix of the RG flow at the fixed point that we call
$\omega_1$ and $\omega_2$. They represent the generalization of the exponent $\omega$ of the $O(N)$
models  and they rule the stability of the fixed point: it is attractive when  $\omega_1$ and $\omega_2$
have both  positive real parts.

\begin{figure}[tbp]
\begin{center}
\begin{picture}(500,175)
\put(95,-6) {(a)}
\put(0,3){\includegraphics[width=0.4\textwidth]{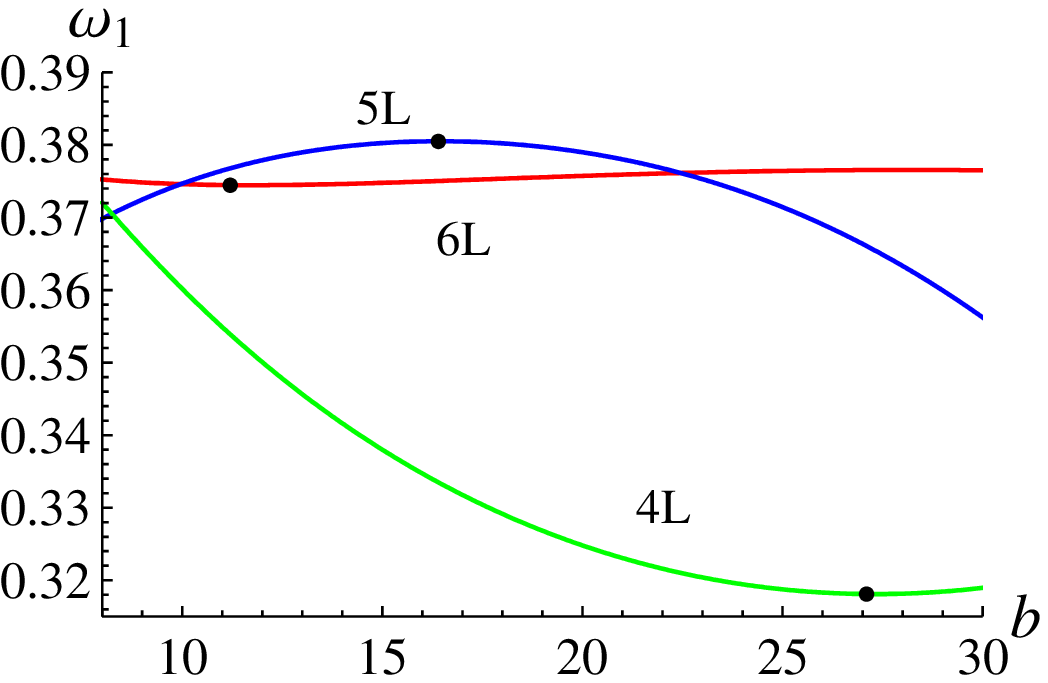}}
\put(315,-6){(b)}
\put(230,3){\includegraphics[width=0.4\textwidth]{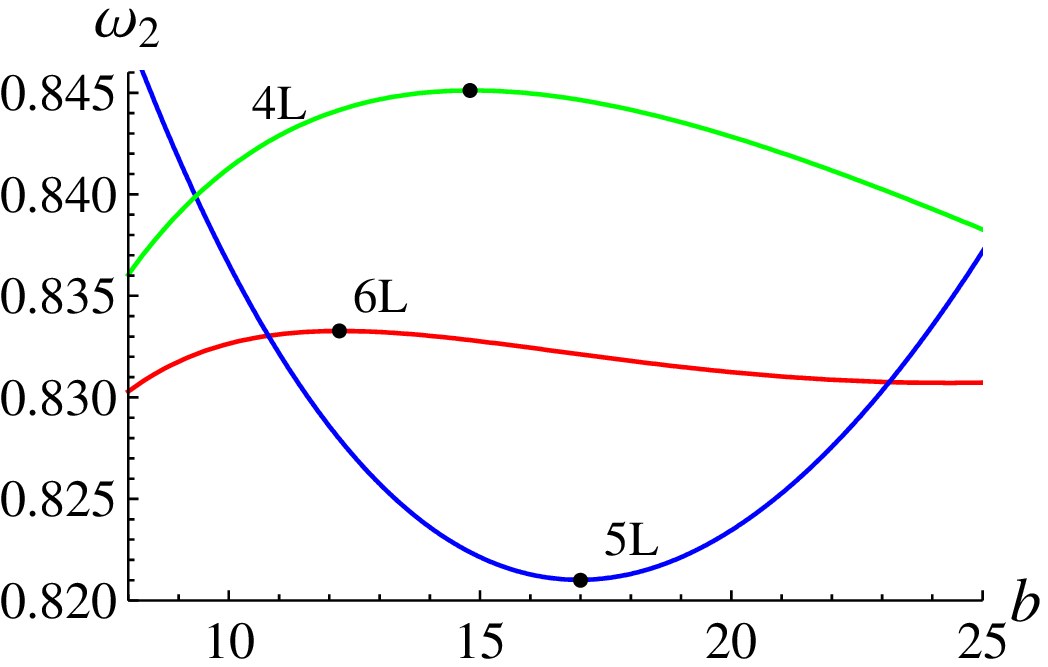}}
\end{picture}
\end{center}
\caption{The exponents $\omega_1$  and $\omega_2$ of the
three-dimensional frustrated model for $N=8$ as  functions of the resummation parameter $b$ 
at  four-, five- and six-loop orders. The dot  on each curve corresponds to a stationary value of $\omega=\omega(\alpha,b)$ in both  $\alpha$ and $b$ directions   with $a$ fixed to its large-order value $a_{\rm lo}\simeq  0.0554$. For $\omega_1$ one has   $(\alpha_{\rm opt},b_{\rm opt})=(6.5,27.1)$ at four loops,  $(\alpha_{\rm opt},b_{\rm opt})=(7.9,16.4)$ at five loops  and
$(\alpha_{\rm opt},b_{\rm opt})=(7.5,11.2)$ at six    loops. For $\omega_2$ one has  $(\alpha_{\rm opt},b_{\rm opt})=(5.1,14.8)$ at four loops,  $(\alpha_{\rm opt},b_{\rm opt})=(7.8,17)$ at five loops  and $(\alpha_{\rm opt},b_{\rm opt})=(7,12.2)$ at six    loops. } 
\label{frN8b}
\end{figure}

We take for $a$ the value  obtained from the large order analysis: $a_{\rm lo}=0.0554$.  We   find that the PMS is satisfied  at  four-, five- and six-loop orders:  for suitable  values of the parameters $b$ and $\alpha$ the two exponents  $\omega_1$ and $\omega_2$  depend weakly on these parameters and are  
reasonably well converged. This is clear from Fig.\ref{frN8b}  where we show the  $b$-dependence of $\omega_1$ and $\omega_2$.  Moreover  the  difference between the values at five and six loops  of, for instance,  $\omega_2$ is small: $\omega_2(L=6)-\omega_2(L=5)\simeq  0.012$. Note that our values of $\omega_1$ and $\omega_2$  in this  case are fully compatible with those obtained in \cite{calabrese03b}.

\begin{figure}[htbp]
\begin{center}
\begin{picture}(500,175)
\put(100,-6) {(a)}
\put(0,3){\includegraphics[width=0.4\textwidth]{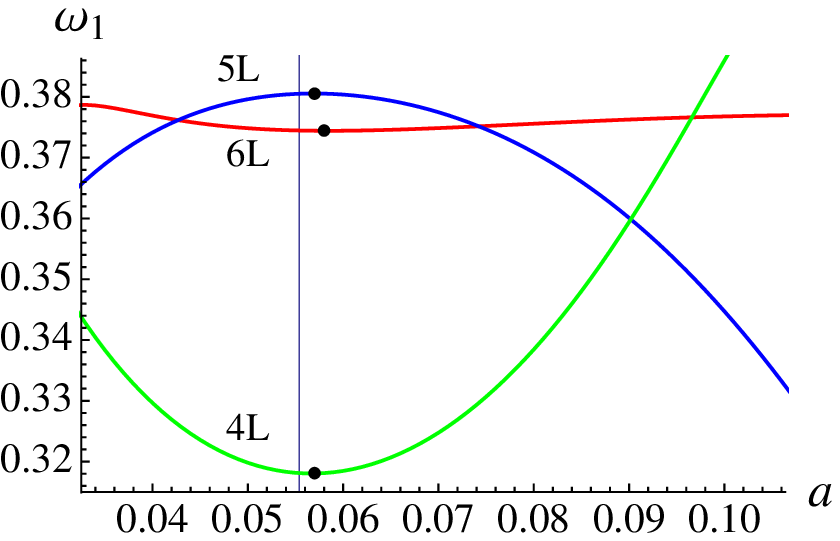}}
\put(310,-6){(b)}
\put(230,3){\includegraphics[width=0.4\textwidth]{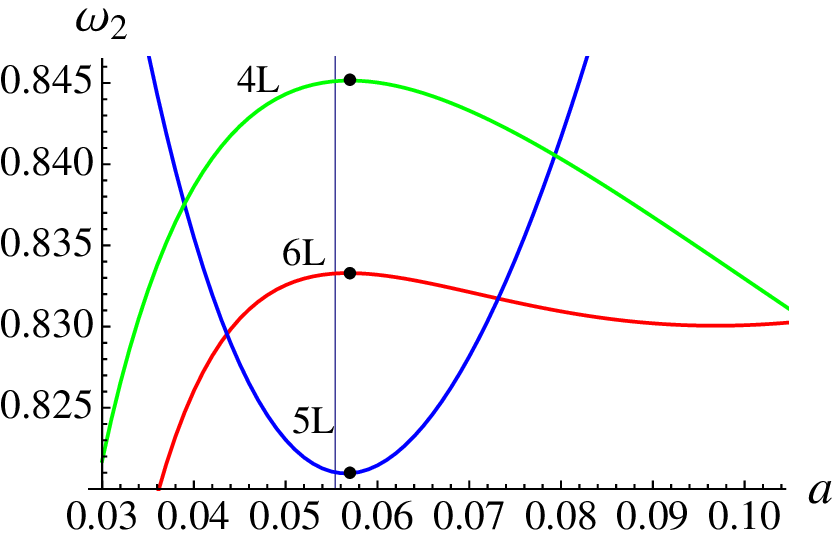}}
\end{picture}
\end{center}

\caption{The exponents $\omega_1$  and $\omega_2$  of the three-dimensional
frustrated model for $N=8$ as functions of the  parameter  $a$ at four-, five- and six-loop orders.  The vertical lines  corresponds to  the  large order value of $a_{\rm lo}=0.0554$. The values chosen for $\alpha$  and $b$  are such that the exponents are at their stationary value when $a = a_{\rm lo}$ (see Fig.\ref{frN8b}).}
\label{frN8a}
\end{figure}

 We have also studied the $a$-dependence of these quantities and find that
the ``optimal'' value of $a$ is close to its large order value ($a_{\rm opt}\simeq a_{\rm lo}=0.0554$)  as it is the case in the $O(N)$ models,  see Fig.\ref{frN8a}. 

These results indicate that  the convergence properties of  the  $N=8$ frustrated model  are globally similar  to those of the $O(N)$ 
 models although  less accurate  very likely  because   in the latter case  the resummation is  less efficient due  to  the presence of two coupling constants.

\subsubsection{The $N=2$ and  $N=3$   frustrated models  in $d=3$ }

We now analyze the physical values of $N$, that is $N=2$ and 3. In \cite{delamotte08}, it has been  found in the 
$\overline{\hbox{MS}}$ scheme that, because 
of the presence of the singularity $S$ (see Fig.\ref{courbes_ncd})  which  exists in this scheme for 
$N\simeq7$ and $d\simeq3.2$, the  results obtained from 
the resummed series above and below $N\simeq 7$ are very different. In the massive scheme, 
the series are  known in integer dimensions only and it is thus  not possible to know whether
this perturbative scheme leads also to the existence of a singularity.  We 
nevertheless show  that, within this scheme,   the results obtained  for $N=2$ and 3  are very different from those obtained 
for $N\ge8$ and are fully compatible with those obtained with the  $\overline{\hbox{MS}}$ scheme \cite{delamotte08}.

\begin{figure}[htbp]
\begin{center}
\begin{picture}(500,175)
\put(92,-5) {(a)}
\put(0,3){\includegraphics[width=0.4\textwidth]{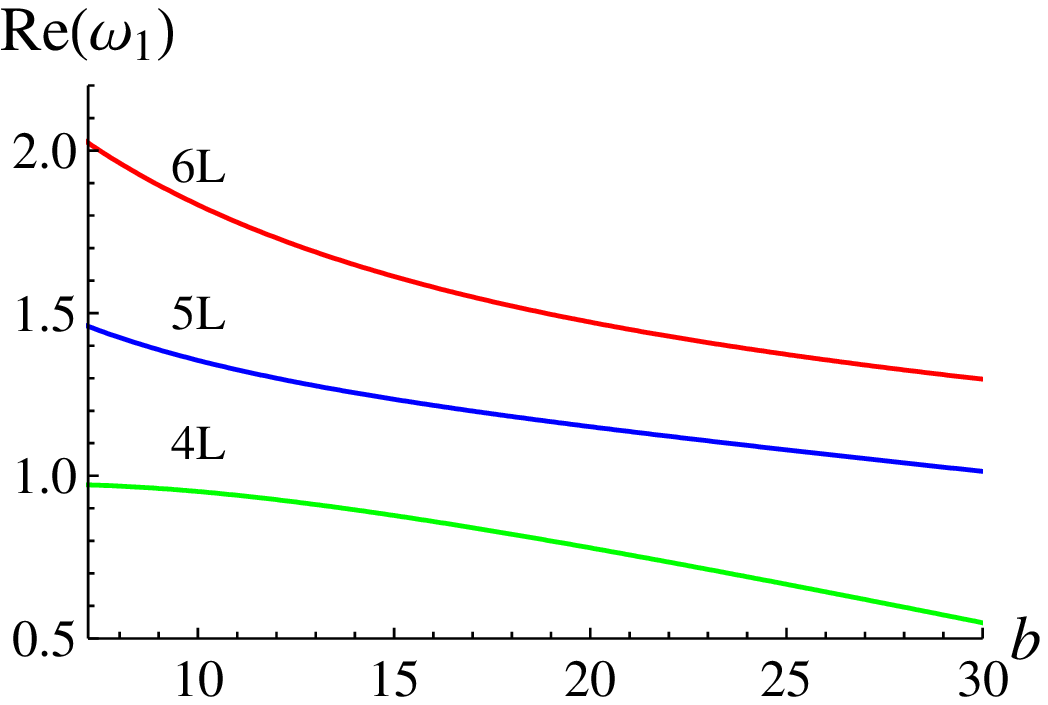}}
\put(320,-5){(b)}
\put(230,3){\includegraphics[width=0.4\textwidth]{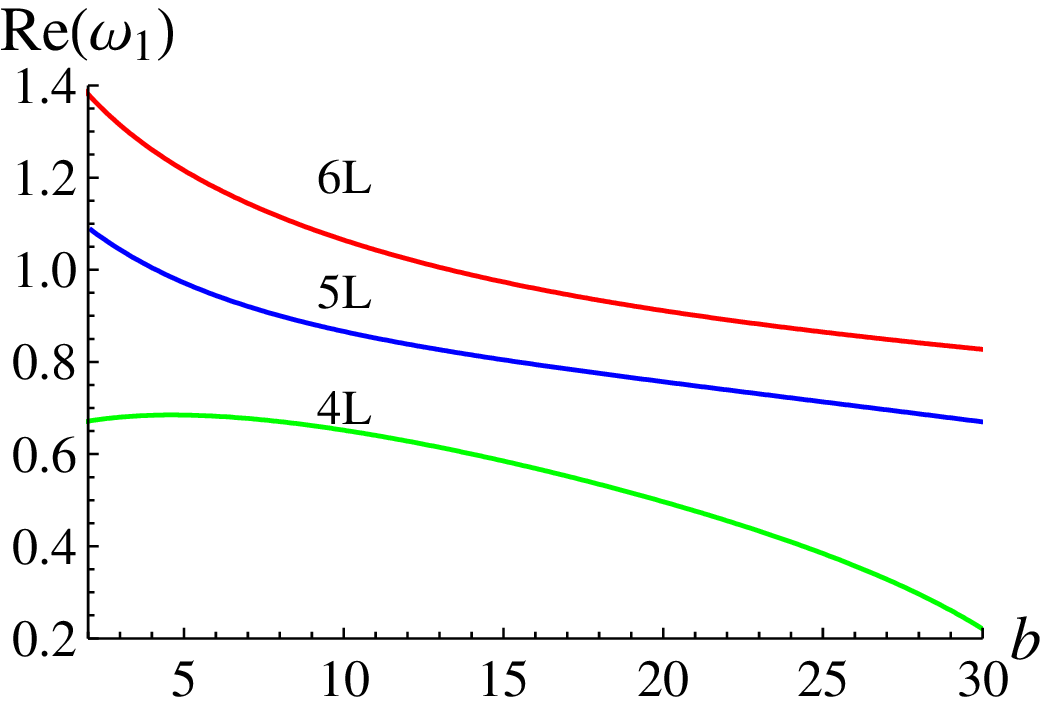}}
\end{picture}
\end{center}
\caption{The (real part of the) exponent $\omega_1$  of the three-dimensional frustrated model   (a) for $N=2$
and  (b) for $N=3$  as a function of  $b$  for $\alpha=6$  at
four, five and six  loops.  The parameter $a$ has been fixed at the value obtained 
from the large order behavior: $a_{\rm lo}=0.1108$ for $N=2$ and $a_{\rm lo}=0.095$ for $N=3$. Other values of the parameter  $\alpha$ give similar results.} 
\label{frN2-3b}
\end{figure}

Let us first recall that for  $N=2$ and $N=3$ the  fixed point is  (in most cases  but not systematically, in particular for values of $\alpha$ different from $1,2,3$) an attractive  focus, that is $\omega_1$ and $\omega_2$ are complex conjugate and  Re($\omega_1$)=Re($\omega_2)>0$. 
We, again, take for $a$ the value 
obtained from the large order analysis: $a_{\rm lo}=0.1108$ for $N=2$ and $a_{\rm lo}=0.095$ for $N=3$.  For these values of $a$ and  for  $L=4,5$ and $6$, we  find that Re($\omega_1$) (or equivalently Re($\omega_2$)) considered as a function of $\alpha$ and
$b$ is {\it   nowhere}   stationary, even approximately, see Fig.\ref{frN2-3b}. Moreover, at fixed $\alpha$ and
$b$, the gap between the values of Re($\omega_1$) at two successive loop-orders: 
Re($\omega_1)(L+1)-$ Re($\omega_1)(L)$, is always large,  of order $0.5$ for $N=2$ and $0.2$ in the $N=3$ case, see Fig.\ref{frN2-3b}. Thus neither the PMS nor the PFAC are  satisfied for these values of $N$.

\begin{figure}[htbp]
\begin{center}
\begin{picture}(500,175)
\put(100,-8) {(a)}
\put(0,3){\includegraphics[width=0.4\textwidth]{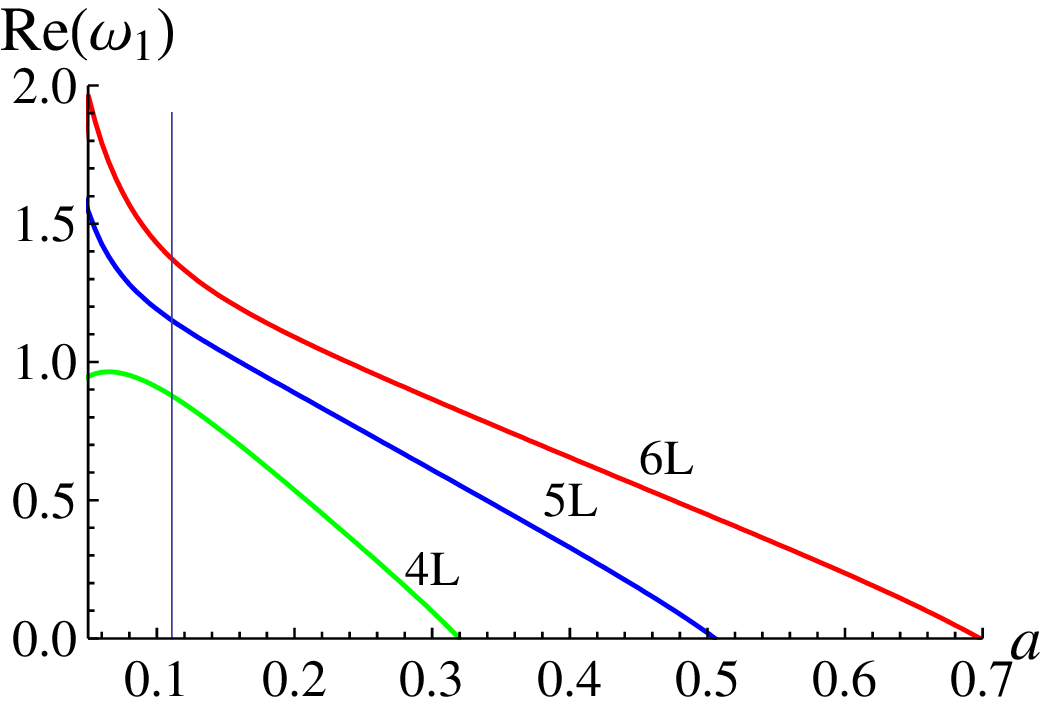}}
\put(325,-8){(b)}
\put(230,0){\includegraphics[width=0.4\textwidth]{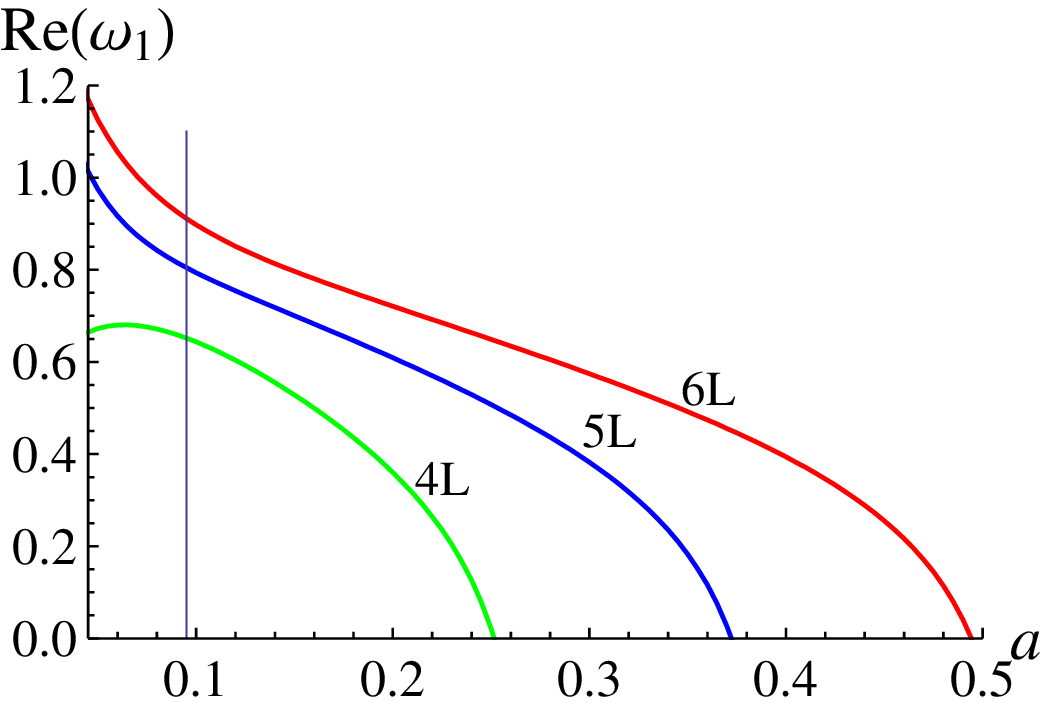}}
\end{picture}
\end{center}
\caption{The (real part of  the)  exponent $\omega_1$ of  the three-dimensional frustrated model  (a)  for  $N=2$
and  (b) for  $N=3$  as a function of  $a$.  The vertical lines  corresponds to $a_{\rm lo}=0.1108$ for $N=2$ and $a_{\rm lo}=0.095$ for $N=3$. One has taken  $\alpha=6$   and
$b=25, 20$ and  15   at  six,  five and  four loops for $N=2$ and $b=20, 15$ and  10  at  six,  five and  four loops for $N=3$. Other values of the parameters $\alpha$ and $b$ give similar results.} 
 \label{frN2-3a}
\end{figure}

 We have also studied the stability of our results with respect to variations
of $a$ for characteristic values of $\alpha$ and $b$, see Fig.\ref{frN2-3a}. Here again we find no stationarity.

From these results it clearly appears that the critical exponents  deduced from the resummed series obtained in the massive scheme in $d=3$  display  
both lack   of  convergence and of stability for the values $N=2$ and 3.  It  seems  therefore very  likely  that the existence of fixed points
for these values of $N$ is an artefact either of the fixed dimension schemes or of the resummation method.
In any case, we confirm by this study that there is no reason coming from the fixed dimension approaches
to question the results obtained either within
the $\epsilon$-expansion or the NPRG and that, very probably,  the transitions found in $d=3$ and $N=2$ and 3
are always of first order.
 
Let us now perform the same kind of analysis for the two-dimensional models.

\subsection{The frustrated models in $d=2$}

As already emphasized, the two-dimensional case is particularly interesting because
of the presence of topological excitations in the Heisenberg case \cite{kawamura84}. Because of the homotopy properties  of the symmetry group $SO(3)$ of these systems, the topological excitations are different from the $O(2)$ vortices encountered in the ferromagnetic XY system. It is still an open question to know whether the deconfinement of these defects could trigger  a  genuine phase transition, as in the Kosterlitz-Thouless case.  Note that such a phenomenon would be surprising since one knows  from the spin-wave -- low-temperature --   approach   \cite{azaria90,azaria92,azaria95}  that, contrary to  the $O(2)$ model,   the  spin-spin correlation length  of $O(3)$ frustrated  models  is  finite at  low  -- but non vanishing  -- temperature and that vortices tend to further disorganize the system.   We  let aside  the  delicate question  of  the very mechanism underlying a  hypothetical genuine  phase transition in these systems  and focus on  the question of the existence of a  finite temperature  fixed point   within  the FD formalism.

As for the  XY  case the  question is to  know whether  there is a unique or two separate (Ising and KT) phase transitions.   From the most recent Monte Carlo  simulations it has  been argued that there are two  distinct but very close phase transitions,  the  Ising  one  taking place  at  the  highest  temperature.  Accordingly one  could expect  the transition to be characterized  by  Ising critical exponents.

In $d=2$ and for the values of $N\ge 4$   there is 
no topological defects. As a consequence, there cannot be any other fixed point but the zero
temperature one.  Thus, for these values of
$N$ and because of Mermin-Wagner's theorem, the correlation length diverges 
 at zero temperature only and with an exponent $\nu$ which is infinite (exponential divergence of 
the correlation length). Moreover  the 
anomalous dimension $\eta$ is always vanishing at a zero temperature fixed point as can be checked on the low temperature 
expansion performed within the non linear sigma model \cite{azaria90,azaria92,azaria95}. Thus, as in the two-dimensional $O(N)$ case,  
any nonvanishing $\eta$   for   $N\ge 4$  must  be considered as an artefact
and, for the perturbation theory, as a signal of an anomalous apparent convergence,   see above.  We start our analysis
by the $N=8$ case and then study the physically  relevant cases: $N=2,3$.

\subsubsection{The $N=8$ case in $d=2$}

\begin{figure}[htbp]
\begin{center}
\begin{picture}(500,125)
\put(0,3){\includegraphics[width=0.35\textwidth]{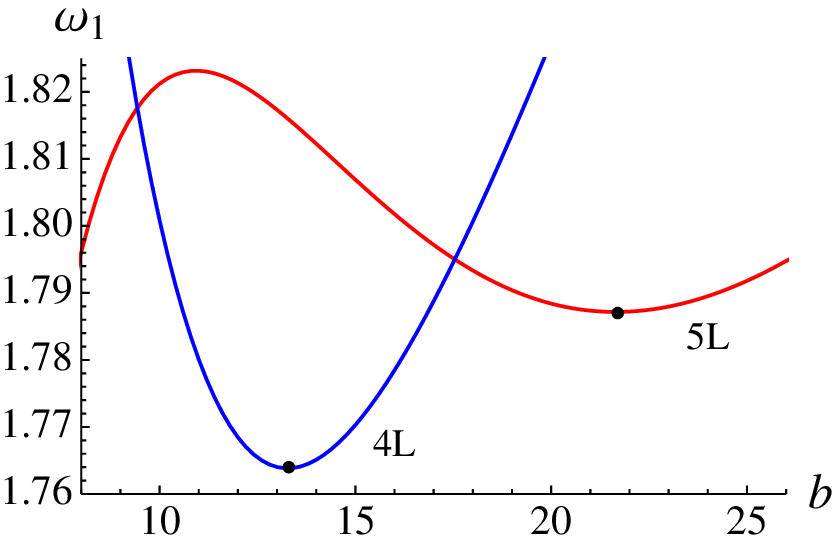}}
\put(90,-10) {(a)}
\put(230,3){\includegraphics[width=0.35\textwidth]{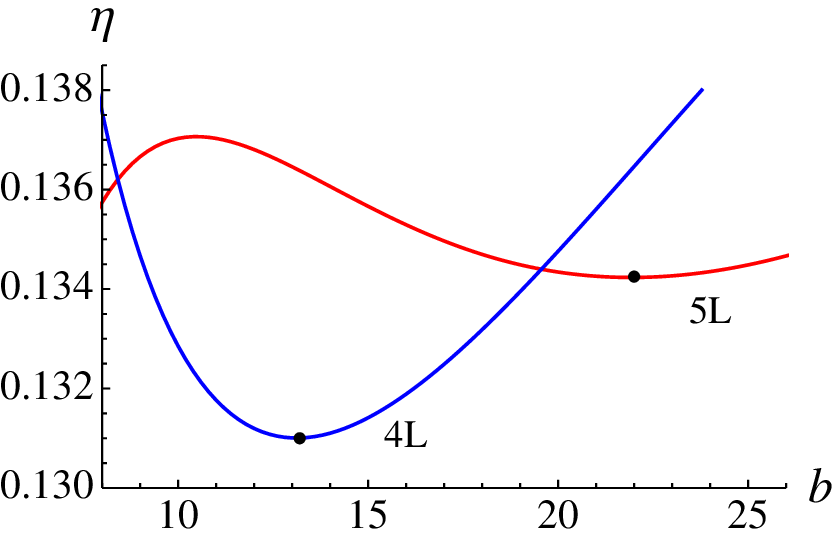}}
\put(325,-10){(b)}
\end{picture}
\end{center}
\caption{The exponents  $\omega_1$ and $\eta$  in the two-dimensional  frustrated $N=8$ case as
a function of  $b$   at four- and five-loop orders.  The dots at each curve corresponds to a 
 stationary value of $\omega=\omega(\alpha,b)$ in both  $\alpha$ and $b$ directions. For  $\omega_1$ one has: $(\alpha_{\rm opt},b_{\rm opt})=(4.7,13.3)$ at  four-loops and  $(\alpha_{\rm opt},b_{\rm opt})=(4.7,21.7)$ at  five loops. For  $\eta$ one has:  $(\alpha_{\rm opt},b_{\rm opt})=(4.55,13.2)$ at four loops and  $(\alpha_{\rm opt},b_{\rm opt})=(4.55,22)$ at five loops.}
\label{eta-omega-frustn8d2l54b}
\end{figure}

For $N=8$ and being given  the previous discussion, one  should  obtain only trivial results as for the critical exponents: $\eta=0$ and $\omega_1=\omega_2=2$.
We have computed  $\eta$ and  $\omega_1$ (the largest eigenvalue)  as functions of the resummation parameters
$\alpha$ and $b$ by taking for $a$ the value computed from the large order behavior: $a_{\rm lo}\simeq0.0895$.
We find a  stationary  solution for the two exponents studied, see 
Figs.\ref{eta-omega-frustn8d2l54b}. However, as in the $O(N)$ models, we find
that the value of $\eta$ thus obtained: $\eta\simeq 0.13$ is unphysical since it should be zero. We find  $\omega_1\simeq 1.79$  which is far from the expected physical value $\omega_1=2$.

\begin{figure}[htbp]
\begin{center}
\begin{picture}(500,125)
\put(150,-10) {(a)}
\put(0,3){\includegraphics[width=0.35\textwidth]{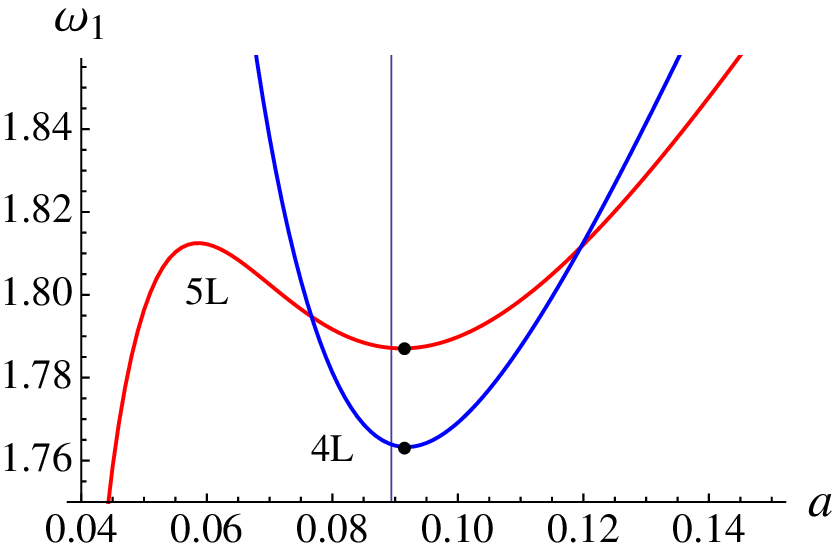}}
\put(375,-10){(b)}
\put(230,3){\includegraphics[width=0.35\textwidth]{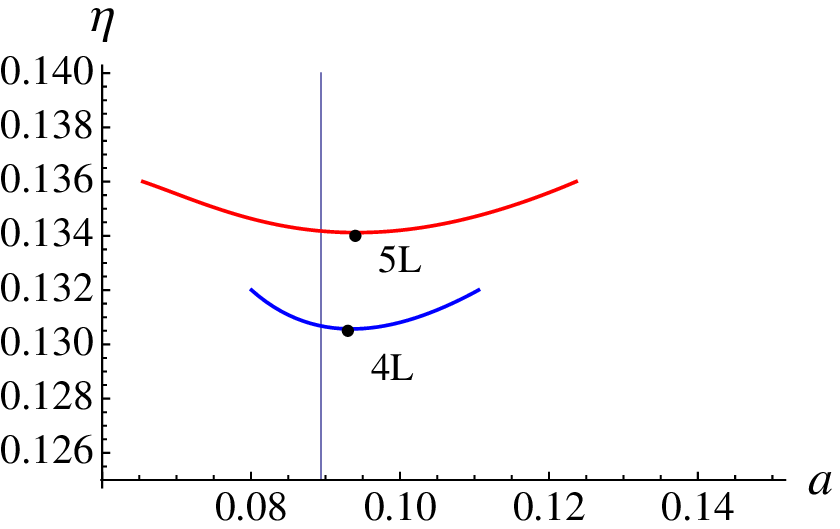}}
\end{picture}
\end{center}
\caption{The exponents $\omega_1$  and $\eta$ in the two-dimensional frustrated   model for $N=8$ as  functions of  $a$  at four and five loops. The vertical lines corresponds to $a=a_{\rm opt}=0.0895$. The values of $\alpha$   and $b$  are such that the exponents are at their stationary  values when $a=a_{\rm lo}$ (see Fig.\ref{eta-omega-frustn8d2l54b}).} 
\label{eta-omega-frustn8d2l54a}
\end{figure}

We have also studied the $a$-dependence of these exponents. Here again, we find
good convergence properties with an extremum  around the value $a_{\rm lo}$, see Fig.\ref{eta-omega-frustn8d2l54a}.

It thus appears  that there  very likely exist in  frustrated models,   as  in   $O(N)$ models,   nonanalytic terms in the $\beta$-functions 
that spoil the convergence of the resummed perturbative expansion of the critical exponents. 
We can already assert that this
 dramatically alters
the relevance of the perturbative $\phi^4$-approach for the study of the two-dimensional frustrated systems.

\subsubsection{The $N=2$ and $N=3$ cases in $d=2$}

 We now perform the same analysis  as above for the physically relevant values of $N$, that is  $N=2$ and 3. Let us first notice that,  for these values of $N$, 
the  fixed point starts to exist  beyond three loop-order only. We fix $a$ at its large
 order value: $a\simeq0.1790$ for $N=2$ and  $a\simeq 0.1534$ for $N=3$.  

\begin{figure}[htbp]
\begin{center}
\begin{picture}(500,125)
\put(85,-10) {(a)}
\put(0,3){\includegraphics[width=0.35\textwidth]{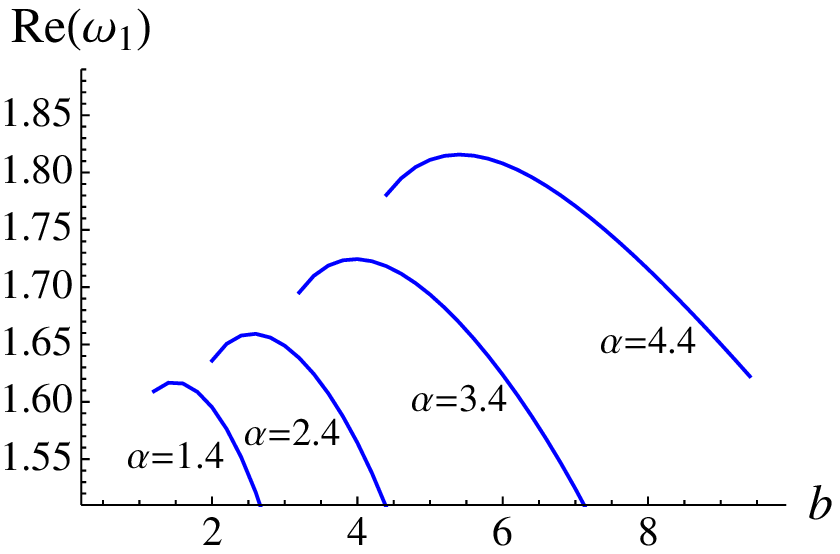}}
\put(300,-10){(b)}
\put(230,3){\includegraphics[width=0.35\textwidth]{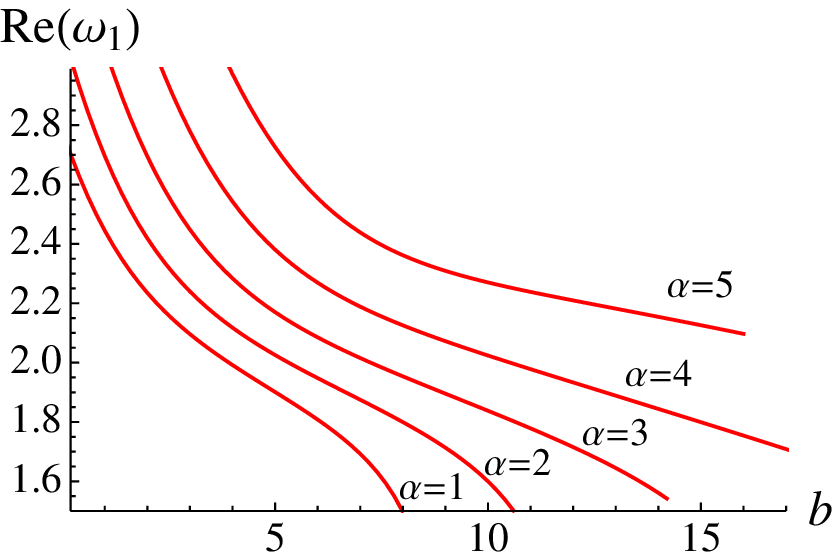}}
\end{picture}
\end{center}
\caption{ The (real part  of the)  exponent $\omega_1$ of the two-dimensional frustrated model for $N=2$
as a function of $b$ for  different values of $\alpha$ (a) at four loops and for $\alpha=1.4, 2.4, 3.4, 4.4 $ (b) at five loops
 and for $\alpha=1, 2, 3, 4, 5 $.  We have chosen  $a\simeq0.1790$. }
\label{omegafrustn2d2alpha}
\end{figure}

Let us first discuss the $N=2$ case.
We find that the correction to scaling exponent $\omega_1$  (and thus $\omega_2$)   is complex for a large range of parameters $\alpha$ and $b$
which means that the fixed point is a focus. We show in Fig.\ref{omegafrustn2d2alpha} that there is  no 
value of $\alpha$ and $b$  where $\omega_1$ is stationary with respect to both parameters.
Moreover, at fixed $\alpha$ and $b$,  the difference between the four- and five-loop results is large. This is a clear signal of the nonconvergence
of the value of $\omega_1$.  In  \cite{calabrese03} an average value for this critical exponent has been proposed:  $\omega_1=2.05(35)\pm i 0.80 (55)$ at five loops. According to our stability and convergence principles this value does not really make sense.

\begin{figure}[h]
\begin{center}
\begin{picture}(500,125)
\put(80,-10) {(a)}
\put(0,3){\includegraphics[width=0.35\textwidth]{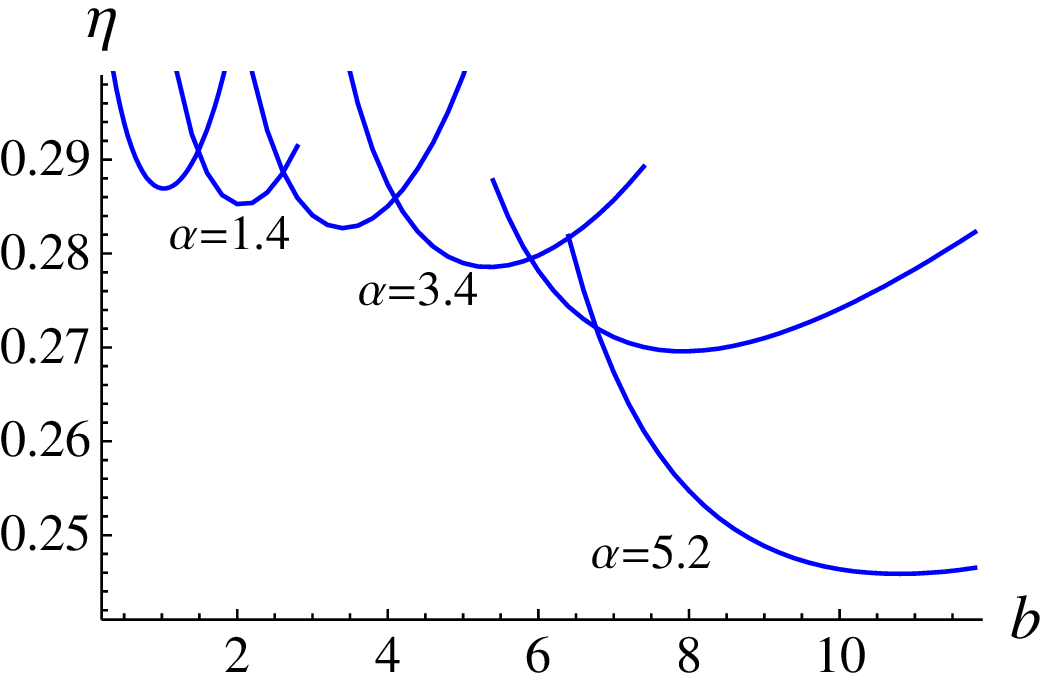}}
\put(325,-10){(b)}
\put(230,3){\includegraphics[width=0.35\textwidth]{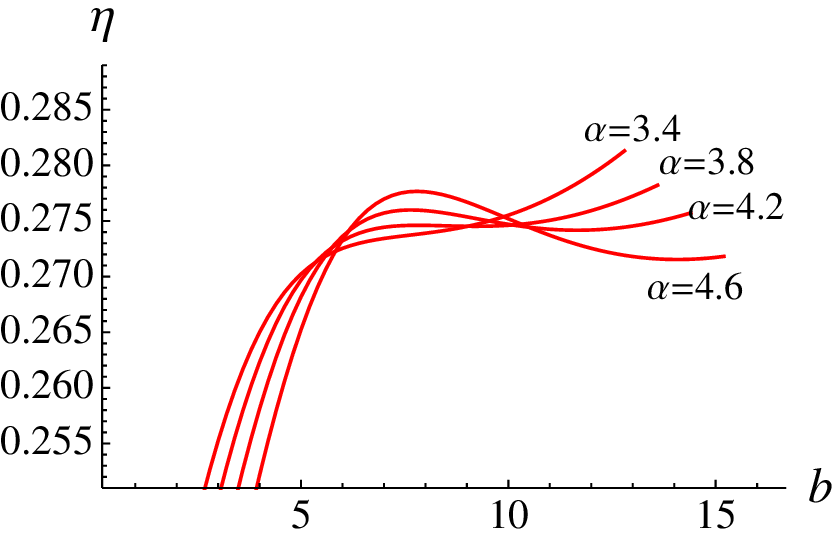}}
\end{picture}
\end{center}
\caption{ The   exponent $\eta$ of the two-dimensional frustrated model for $N=2$
as a function of $b$ for  different values of $\alpha$.  (a) at four loops and for $\alpha=0.3, 1.4, 2.4, 3.4, 4.4, 5.2 $ (b) at five loops
 and for $\alpha=3.4, 3.8, 4.2, 4.6$. We have chosen  $a\simeq0.1790$.}
\label{etafrustn2d2alpha}
\end{figure}

The situation is a little bit different for the exponent $\eta$. At four loops  $\eta$ is nowhere stationary in the 
$\alpha$-direction as can be seen on Fig.\ref{etafrustn2d2alpha}-a)  whereas there is an almost stationary value  $\eta\simeq0.275$   at five loops  in both $\alpha$ and $b$ directions for $\alpha\simeq4.2$ and $b\simeq11$, see Fig.\ref{etafrustn2d2alpha}-b) (which is compatible with the value  given in \cite{calabrese03}  where $\eta=0.28(8)$). We have performed the analysis of the stability of our results for $\eta$ when $a$ is varied around  $a_{\rm lo}$ at fixed $\alpha$ and $b$.  We find that indeed the five-loop results do not vary much with $a$ and that the optimal value of $a$ is close to $a_{\rm lo}$.

We conclude that the results for the $N=2$ case show no convergence with  the loop order  and a poor stability with respect to variations of  $\alpha$ and $b$   but perhaps for the exponent $\eta$ at five loops.  Let us notice that the value found $\eta\simeq 0.275$ is relatively close to the exact value expected for an Ising transition ($\eta=0.25$). However, at the same time, it is  far  from  the five-loop value  $\eta=0.146$ \cite{orlov00} obtained directly with  the $\pmb \phi^4$  field theory.  We shall  develop on this below.

\begin{figure}[htbp]
\begin{center}
\begin{picture}(500,125)
\put(80,-10) {(a)}
\put(0,3){\includegraphics[width=0.35\textwidth]{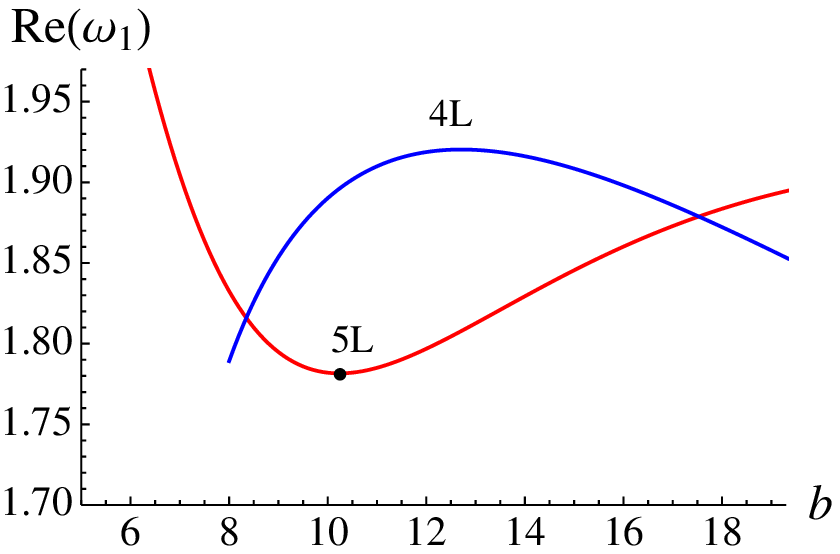}}
\put(305,-10){(b)}
\put(230,3){\includegraphics[width=0.35\textwidth]{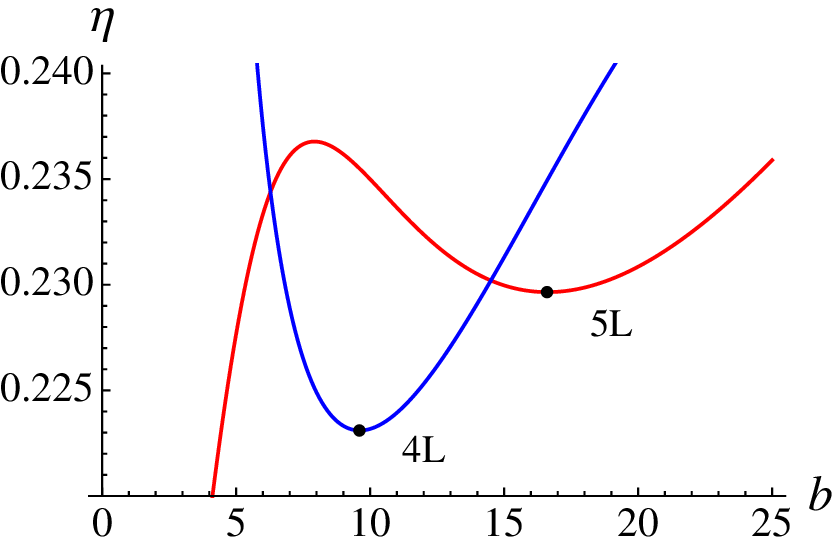}}
\end{picture}
\end{center}
\caption{The  (real part of the) exponent $\omega_1$ and $\eta$ of the two-dimensional frustrated model for $N=3$
as  functions of $b$   at  four and five loops.   (a) For $\omega_1$ there is a stationary point only at five loops  for $(\alpha_{\rm opt},b_{\rm opt})=(5.95,10.1)$  (b) For $\eta$   there are  stationary points at four loops at $(\alpha_{\rm opt},b_{\rm opt})=(4.28,9.6)$  and at five loops at $(\alpha_{\rm opt},b_{\rm opt})=(4.45,16.6)$.  The value of $a$ has been taken equal to its large order  value $a\simeq0.1534$.}
\label{etafrustn3d2alpha}
\end{figure}

Let us now examine the $N=3$ case, Fig.\ref{etafrustn3d2alpha}.  The fixed point is again  a focus. The difference with the $N=2$ case  is that there now exists a  stationary point for Re$(\omega_1)$  at five loops  for  $\alpha\simeq  5.95$ and $b=10.25$, but not at four  loop-order, see Fig.\ref{etafrustn3d2alpha}-a), where there  is no stationarity w.r.t. $\alpha$.  At this stationary point one has Re$(\omega_1)\simeq 1.78$ (which is compatible with the result found in \cite{calabrese03}: Re$(\omega_1)=1.55(25)$ that anyway displays a large error bar).  We find stationary points  for $\eta$ at four-  and five-loop orders,  see Fig.\ref{etafrustn3d2alpha}-b). At five loops the value of $\eta$ at the stationary point is $\eta=0.23$ (that compares  well with  the value $\eta=0.23(5)$ of \cite{calabrese03}).  The convergence seems better in this $N=3$ case than in the $N=2$ case since  now  both $\omega_1$ and $\eta$  display stationary values.

\begin{figure}[htbp]
\begin{center}
{\includegraphics[width=0.4\textwidth]{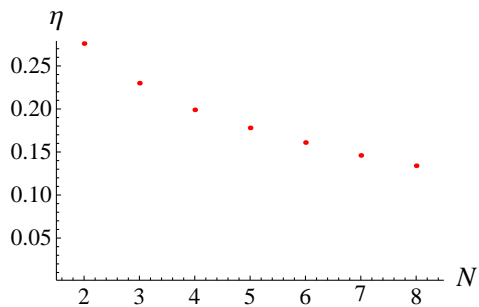}}
\end{center}
\caption{Exponent $\eta$ of the two-dimensional frustrated model as a
function of the number of spin components $N$.}
\label{plotetaN}
\end{figure}

 From the discussion above one could be tempted   to  conclude,  in the $N=3$ case, that the value $\eta=0.23$, although affected by a large error bar  ($\delta \eta=0.05$ according to  \cite{calabrese03}),   is sufficiently large  to  ensure  that $\eta$  does not vanish,  as claimed in  \cite{calabrese03}.   In this case  the transition would  be non-trivial,  that is,  would occur at finite temperature.   We now argue that  the results obtained at five loops  are not sufficiently accurate to support this conclusion. The reason is that  the error on $\eta$ is, in fact,  underestimated. To  see this we have computed  $\eta(N)$ (according to our two principles)  for all values of  $N$ between 2 and 8, see Fig.\ref{plotetaN}.  As already emphasized there  cannot exist nontrivial fixed points  and, thus,  nonvanishing  anomalous dimensions  $\eta$ for  any   value of $N\ge 4$.  As seen on Fig.\ref{plotetaN} this  is   violated by the perturbative results at five loops. This implies  that the error $\delta\eta$ on $\eta$ at five loops  is of order $\eta$ itself, that is  in the $N=4$ case,  of order  $0.20$.    Being given that $\eta(N)$ is   monotonically decreasing,  the error bar  increases as $N$ decreases.  In the $N=3$ case,   the error bar is thus   at least equal to $0.20$  and since $\eta(N=3)$ is found to be  equal to $0.23$  it  is impossible to  conclude that $\eta$ is nonvanishing in this case. While  our  considerations extend also very likely  to the $N=2$ case as for the existence of a large error on the result, this case is  particular. Indeed  for $N=2$  one expects  $\eta=0.25$ since the transition likely  belongs  to the Ising universality class. At first sight the value found at five loops ($\eta=0.275$) could seem encouraging.  However let us recall that one finds $\eta=0.146$  from the one-component  $\phi^4$ model in $d=2$    \cite{orlov00,pogorelov07} which is very far from the expected result.   As can be  seen from the $N\ge 4$ results,  the series for frustrated magnets do not  exhibit  better convergence properties than those of the $\phi^4$ model and  thus the value of $\eta$  found in the $N=2$ frustrated case should very likely  be  interpreted as  a numerical coincidence.  This conclusion is reinforced by the fact that, as explained previously,  the  stability properties of  $\eta$  in the $N=2$ case  are also  unsatisfactory, see Fig.\ref{etafrustn2d2alpha}.

\section{Conclusion}

We have investigated the series obtained  from FD perturbative approaches of $O(N)$ models and frustrated magnets both in $d=2$ and in $d=3$ at five-    and six-loop orders  respectively.  From a general point of view the result of our study is that only the $O(N)$ models  in $d=3$ provides unambiguous and precise results.  For frustrated magnets, our results in $d=3$, that  show an absence  of stationarity of the exponents considered as  functions of  the resummation parameters $\alpha$ and $b$  and  a bad  convergence with the number of loops, provide strong support  to  the spurious character of   the fixed points  found  for $N=2$ and  $N=3$.   Without providing  a definitive answer to the question of the nature of the phase transition that   frustrated magnets undergo in $d=3$  our  results  weaken a lot the predictions of a second order behaviour. Since all other   studies than the FD approaches ($\epsilon$-expansion, NPRG) predict  first order behaviour  (see \cite{delamotte03})    we are naturally  led to the conclusion that   three-dimensional frustrated magnets  should  always  exhibit  first order behaviours. 

  In  $d=2$,  the situation is more ambiguous  since the critical  exponents satisfy the PMS  in some cases  (at some orders and for some critical exponents).    At first sight,  one could deduce from  these results the existence, for Heisenberg spins,  of a finite temperature phase transition triggered by the deconfinement of topological excitations.  However a careful comparative study between the $O(N)$ and frustrated models shows  that the presence of nonanalycities  spoils  the determination of the critical exponents and forbids  to  conclude. 
  
   It is not clear  that  only a few higher orders of the perturbative expansion  would be sufficient to clarify the situation and one has to think  about another approach in both $d=2$ and $d=3$. From this point of view  the NPRG  seems to be able to circumvent the main difficulties. Indeed being not based on  a perturbative expansion (in the traditional sense) it  does not suffer  from   some of the problems encountered in the  weak coupling approaches.  In particular it  seems to  be unaffected by the problems of nonanalyticities since   the value found for $\eta$ in the $d=2$  Ising case is, within this approach,  found equal to $0.254$ \cite{benitez09}, in excellent agreement with the exact result.   The  $d=2$  case for both  frustrated Heisenberg and XY systems  is under investigation \cite{delamotte10}.

Let us  finally   emphasize that  the methodology put forward in this article  could be  relevant  for any system analyzed  within the FD perturbative approach.   
Indeed, in this case,  the existence of  spurious  fixed points  is the generic case and one has to be especially careful when  fixed points that have no counterpart  within the $\epsilon$-expansion approach occur.   In such circumstances,  the  principles employed here   -- PMS and PFAC --  can be  of  great  interest  to reject or to accept the fixed points  as physical solutions.

\acknowledgements{We wish to acknowledge the CNRS-NAS Franco-Ukrainian bilateral exchange program, project  `Critical behavior of frustrated systems'. This work was supported in part by the Austrian Fonds zur F\"orderung der wisserschaftlichen Forschung under project No. P19583-N20 (Yu. H.). M.D. wishes to thank the grant of National Academy of Sciences of Ukraine for  young scientists.}

\end{document}